\documentclass[lettersize,journal]{IEEEtran}

\usepackage{amsmath}
\usepackage{amssymb}
\usepackage{mathtools}
\usepackage{bm}
\usepackage{graphicx}
\usepackage{caption}
\usepackage{subcaption}
\usepackage{cite}
\usepackage{tikz}
\usetikzlibrary{spath3,intersections,positioning,calc,decorations.pathreplacing,patterns,decorations.pathmorphing,decorations.markings}
\usepackage[percent]{overpic}
\usepackage{siunitx}
\usepackage[hidelinks]{hyperref}

\DeclareMathAlphabet{\mbf}{OT1}{ptm}{b}{n}
\newcommand{\mc}[1]{\ensuremath{\mathcal{#1}}}
\newcommand{\mbs}[1]{\ensuremath{\boldsymbol{#1}}}
\newcommand{\trans}{{\ensuremath{\mathsf{T}}}}
\newcommand{\herm}{{\ensuremath{\mathsf{H}}}}
\newcommand{\frob}{{\ensuremath{\mathsf{F}}}}

\newcommand{\trace}{{\ensuremath{\mathrm{tr}}}}

\newcommand{\Hinf}{$\mathcal{H}_\infty$}
\newcommand{\Htwo}{$\mc{H}_2$}

\newcommand{\mbfhat}[1]{\ensuremath{\hat{\mbf{#1}}}}

\newcommand{\bma}[1]{\left[\begin{array}{#1}}
\newcommand{\ema}{\end{array}\right]}

\begin{document}

\tikzstyle{block_ob} = [
draw,
on grid,
minimum width=1.2cm,
minimum height=1.2cm
]

\tikzstyle{bdarrow} = [
-stealth,
]

\tikzstyle{sumt} = [
draw,
on grid,
circle,
minimum size=0.6cm,
]

\tikzstyle{diff_ns} = [
sumt,
label=80:{\small $+$},
label=260:{\small $-$}
]

\tikzstyle{diff_nw} = [
sumt,
label=170:{\small $+$},
label=80:{\small $-$}
]

\tikzstyle{sum_nw} = [
sumt,
label=170:{\small $+$},
label=80:{\small $+$}
]

\tikzstyle{sum_ne} = [
sumt,
label=350:{\small $+$},
label=80:{\small $+$}
]

\begin{minipage}{\textwidth}
    \centering
    This work has been submitted to the IEEE for possible publication. Copyright
    may be transferred without notice, after which this version may no longer be
    accessible.
\end{minipage}
\pagenumbering{gobble}
\clearpage
\pagenumbering{arabic}

\title{Uncertainty Modelling and Robust Observer Synthesis using the Koopman Operator}

\author{%
    Steven~Dahdah,~\IEEEmembership{Student~Member,~IEEE,}
    James~Richard~Forbes,~\IEEEmembership{Member,~IEEE}%
    \thanks{This work is supported by the Natural Sciences and Engineering
        Research Council of Canada (NSERC) Discovery Grants program,
        the \textit{Fonds de recherche du Qu{\'e}bec --- Nature et technologies}
        (FRQNT), the \textit{Institut de valorisation des donn{\'e}es} (IVADO), the
        Canadian Institute for Advanced Research (CIFAR), and the \textit{Centre de
            recherches math{\'e}matiques} (CRM), as well as by Mecademic through the
        Mitacs Accelerate program.}%
    \thanks{S. Dahdah and J. R. Forbes are with the
        Department of Mechanical Engineering, McGill University, Montreal QC
        H3A~0C3, Canada (e-mail:
        \href{mailto:steven.dahdah@mail.mcgill.ca}{steven.dahdah@mail.mcgill.ca};
        \href{mailto:james.richard.forbes@mcgill.ca}{james.richard.forbes@mcgill.ca}).}%
}

\markboth{}{Dahdah and Forbes: Uncertainty Modelling and Robust Observer Synthesis using the Koopman Operator}

\IEEEpubid{}

\maketitle

\begin{abstract}
    This paper proposes a robust nonlinear observer synthesis method for a
    population of systems modelled using the Koopman operator. The Koopman
    operator allows nonlinear systems to be rewritten as infinite-dimensional
    linear systems. A finite-dimensional approximation of the Koopman operator
    can be identified directly from data, yielding an approximately linear
    model of a nonlinear system.
    The proposed observer synthesis method is made possible by this linearity
    that in turn allows uncertainty within a population of Koopman models to be
    quantified in the frequency domain. Using this uncertainty model, linear
    robust control techniques are used to synthesize robust nonlinear Koopman
    observers.
    A population of several dozen motor drives is used to experimentally
    demonstrate the proposed method. Manufacturing variation is characterized in
    the frequency domain, and a robust Koopman observer is synthesized using
    mixed \Htwo{}-\Hinf{}~optimal control.
\end{abstract}


\section{Introduction}
\IEEEPARstart{T}{he} design of controllers and observers that are robust to
plant uncertainty or variation is a challenging engineering problem,
particularly when the plant is nonlinear.
A robust controller or observer provides performance guarantees when deployed on
any plant within a specified uncertainty set.
This uncertainty set could represent a population of nominally identical systems
with some variability, such as a product subject to manufacturing variation, but
it could also represent the effect of modelling error or neglected dynamics for
a single system.
By explicitly considering uncertainty in the design process, robust controllers
and observers can improve yield, reliability, and safety in real-world
scenarios.

Robust control theory, which can be used to synthesize either controllers or
observers, is well-developed but generally limited to linear
plants~\cite{skogestad_2006_multivariable, green_linear_1994, zhou_robust_1995}.
In this framework, nonlinear effects are typically treated as perturbations to a
linear nominal plant~\cite[\S9.1]{zhou_robust_1995}.
One approach to extending robust control theory to nonlinear plants involves the
Koopman operator~\cite{koopman_hamiltonian_1931, mezic_2019_spectrum,
    budisic_applied_2012, mauroy_2020_koopman}, a tool that allows nonlinear systems
to be represented globally by infinite-dimensional linear systems. This is
achieved by viewing their dynamics in terms of a set of nonlinear lifting
functions whose time evolution is governed by the Koopman operator.
Finite-dimensional approximations of the Koopman operator can be identified from
data and used in the robust control framework with little modification.

In this paper, a robust nonlinear observer design methodology based on a
population of approximate Koopman models identified using input-output data is
presented.
Then, an industrially relevant nonlinear observer synthesis example is
discussed, where experimental data from a batch of 38 motor drives with Harmonic
Drive gearboxes is used to demonstrate the proposed method.
Harmonic Drive gearboxes are common in aerospace, robotics, and
industry due to their compact form factor, high reduction ratio, and lack of
backlash~\cite{ghorbel_1998_kinematic, tuttle_1996_nonlinear}.
However, these gearboxes are affected by nonlinear oscillations that can degrade
tracking performance and excite vibration modes in the systems where they are
used~\cite{ghorbel_1998_kinematic, tuttle_1996_nonlinear}.
Motor drives with Harmonic Drive gearboxes therefore require nonlinear nominal
models to synthesize a robust observer that can predict this behaviour.

\subsection{Related work}
Observer synthesis methods for specific types of uncertain nonlinear systems
have previously been presented in the literature.
Gain-scheduled linear observers are designed in~\cite{wang_2000_a,
marquez_2005_robust_state_observer, etienne_2022_robust} for linear and bilinear
parameter varying systems.
\Hinf{}~optimal observers for systems with Lipschitz nonlinearities are
synthesized in~\cite{lu_2004_robust, pertew_2005_synthesis,
abbaszadeh_2008_robust, abbaszadeh_2008_lmi}.
A sliding-mode \Hinf{}~optimal observer for Lipschitz nonlinear systems is
proposed in~\cite{raoufi_2010_sliding}.

Early work using the Koopman operator for state estimation introduces the
Koopman observer form (KOF), a particular state-space form computed from the
estimated eigendecomposition of the Koopman operator~\cite{surana_2016_koopman,
surana_2016_linear}.
\IEEEpubidadjcol{}
The KOF is used in~\cite{gomez_2019_data-driven} to design a Koopman Kalman
filter to estimate the flow field near an actuated airfoil from pressure
measurements.
A generalized maximum likelihood variant of the Koopman Kalman filter is
introduced in~\cite{netto_2018_a}, where it is used to estimate the rotor angle
and speed of a series of synchronous generators.
In~\cite{lee_data-driven_2024}, a Koopman Luenberger observer is designed to
detect faulty actuators in a multirotor system.
Koopman observer design for models with bilinear lifting functions is considered
in~\cite{lambe_2024_on}.
Koopman state estimation in a batch optimization framework is discussed
in~\cite{guo_2022_koopman, guo_2024_data-driven}, while simultaneous
localization and mapping (SLAM) is additionally considered
in~\cite{guo_2024_data-driven}.

To date, robust observer synthesis with the Koopman operator has not been
considered, but robust linear time-invariant (LTI) controller synthesis using
the Koopman operator has been addressed in~\cite{uchida_2021_data-driven,
strasser_2023_robust, strasser_2024_koopman-based, tianyi_2024_dual-loop,
eyuboglu_2024_data-driven}.
In~\cite{uchida_2021_data-driven}, \Htwo{}~optimal control is used to synthesize
controllers that perform robustly in the face of uncertainty modelled by
polytope sets of Koopman models. The uncertainty is due to the use of multiple
datasets to identify multiple Koopman models of the same system. The synthesized
controllers are tested in simulation using a Duffing oscillator system and the
Kroteweg--De~Vries partial differential equation.
Uncertainty due to the finite-dimensional approximation of an
infinite-dimensional Koopman operator is represented
in~\cite{strasser_2023_robust, strasser_2024_koopman-based}. Controllers are
designed to guarantee stability in the largest possible regions of attraction
and are tested on simulated Van~der~Pol oscillator~\cite{strasser_2023_robust}
and inverted pendulum~\cite{strasser_2024_koopman-based} systems.
Biased Koopman models identified from noisy data are considered
in~\cite{tianyi_2024_dual-loop}, where uncertainty is modelled based on a known
sector bounded model mismatch. A dual-loop \Hinf{}~control approach is used to
attain both robust stability and robust performance guarantees. The approach is
demonstrated using a simulated Van~der~Pol oscillator.
In~\cite{eyuboglu_2024_data-driven}, robust \Hinf{}~control is applied to a
linear parameter-varying Koopman representation. Uncertainty due to the linear
parameter-varying nature of the model, along with known model approximation
error, is modelled in the frequency domain. The uncertainty is bounded using
constant weighting functions in an additive uncertainty representation. A mixed
\Htwo{}-\Hinf{} control problem is solved to guarantee the robust stability of
the Koopman controller. The performance of the synthesized controller is
demonstrated on a simulated bilinear motor model.
While robustness is not considered in~\cite{ganz_2021_data-driven}, a
sophisticated \Hinf{}~design procedure with dynamic performance weighting
functions is outlined in the lifted space. The synthesized \Hinf{}~controller is
demonstrated on simulated Van~der~Pol oscillator and two-mass-Duffing-spring
systems.
So far, no Koopman control approaches have been demonstrated outside of
simulation, and no Koopman uncertainty models have characterized variation
within a population of nominally identical plants subject to individual
variation.

For the sake of completeness, note that comprehensive reviews of Koopman control
approaches based on the linear-quadratic regulator (LQR) and model predictive
control (MPC) can be found in~\cite{mauroy_2020_koopman, brunton_2022_modern,
otto_2021_koopman}.
Koopman-based MPC algorithms that consider other definitions of robustness
include~\cite{mamakoukas_2022_robust, zhang_2022_robust,
gholaminejad_2023_stable, thomas_2024_koopman}.

\subsection{Contribution}
The key methodological contributions of this paper are
\begin{enumerate}
    \item the frequency-domain quantification of uncertainty in a population of
          Koopman models, and
    \item a robust nonlinear observer synthesis approach for a population of
          Koopman models.
\end{enumerate}
Furthermore, this work extends the input-output observer framework
of~\cite{marquez_2003_a_frequency_domain, marquez_2005_robust_state_observer},
by introducing mixed \Htwo{}-\Hinf{}~optimal control and non-additive
uncertainty forms to the method.

The proposed observer synthesis method is demonstrated experimentally on a
population of motor drives, leading to the following technological
contributions.
\begin{enumerate}
    \item The creation of a publicly available dataset consisting of
        trajectories from 38 individual motor drives in loaded and unloaded
        conditions~\cite{dahdah_2024_quantifying}.
    \item The development of physics-inspired Koopman lifting functions 
        for motor drives with Harmonic Drive gearboxes.
    \item The evaluation of multiple different uncertainty forms for each choice
        of nominal plant model for a population of Koopman models.
    \item The use of nontrivial weighting functions to bound residuals for a
          population of Koopman models.
    \item The experimental, rather than simulated, validation of a robust
          nonlinear observer based on mixed \Htwo{}-\Hinf{}~optimal control.
\end{enumerate}
Two additional minor contributions are discussed in the Appendices. The first is
an outlier detection approach for incorrectly-installed motor drives using the
frequency-domain uncertainty model, and the second is a phase offset calibration
procedure for Harmonic Drive gearbox oscillations.

The aforementioned methodological contributions are general, and can be applied
to a wide range of systems, while the technological contributions are related to
specific challenges posed by the motor drive system under consideration.

\subsection{Outline}
The remainder of this paper is structured as follows.
Section~\ref{sec:background} summarizes robust control theory, its applications
to robust observer synthesis, and Koopman operator theory.
Section~\ref{sec:methodology} describes the proposed robust nonlinear observer
synthesis methodology. Section~\ref{sec:experimental} presents experimental
validation of the methodology using a dataset of 38 nonlinear motor drives.
Finally, Section~\ref{sec:conclusion} concludes the paper.

Appendix~\ref{sec:app_outlier} discusses how frequency-domain uncertainty
modelling can be used to detect outlier motor drives, and
Appendix~\ref{sec:app_calibration} presents a phase offset calibration procedure
for motor drives with Harmonic Drive gearbox oscillations.

\section{Background}\label{sec:background}

\subsection{Robust control theory}
This section outlines the robust control fundamentals required to quantify
uncertainty within a population of LTI systems and synthesize a robust
controller or observer using that uncertainty model.
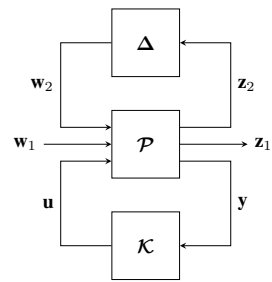
\begin{figure}[htbp]
    \centering
    \scalebox{0.75}{\begin{tikzpicture}
        \node[
            block_ob,
        ] (P) at (0, 0) {$\mbs{\mc{P}}$};
        \node[
            block_ob,
            below = 1.8cm of P,
        ] (K) {$\mbs{\mc{K}}$};
        \node[
            block_ob,
            above = 1.8cm of P,
        ] (Delta) {$\mbs{\Delta}$};
        \draw[bdarrow] ([yshift=-0.3cm]P.east) -- ++(0.9cm, 0cm) |- (K.east)
            node[pos=0.25, right] {$\mbf{y}$};
        \draw[bdarrow] (K.west) -- ++(-0.9cm, 0cm) |- ([yshift=-0.3cm]P.west)
            node[pos=0.25, left] {$\mbf{u}$};
        \draw[bdarrow] ([yshift=+0.3cm]P.east) -- ++(0.9cm, 0cm) |- (Delta.east)
            node[pos=0.25, right] {$\mbf{z}_2$};
        \draw[bdarrow] (Delta.west) -- ++(-0.9cm, 0cm) |- ([yshift=0.3cm]P.west)
            node[pos=0.25, left] {$\mbf{w}_2$};
        \draw[bdarrow] (P.east) -- ++(1.2cm, 0cm)
            node[pos=1, right] {$\mbf{z}_1$};
        \draw[bdarrow] (P.west) ++(-1.2cm, 0cm) -- (P.west)
            node[pos=0, left] {$\mbf{w}_1$};
    \end{tikzpicture}}
    \caption{%
    The generalized plant $\mbs{\mathcal{P}}$ for the robust optimal
    control problem in feedback with a controller $\mbs{\mathcal{K}}$ and an
    uncertainty block $\mbs{\Delta}$.}\label{fig:generalized_plant_delta}
\end{figure}

\subsubsection{Notation}
Let $\mbf{y}: \mathbb{Z}_{\geq 0} \to \mathbb{R}^{n \times 1}$ denote a
discrete-time signal whose value at time $k$ is
$\mbf{y}_k$~\cite[\S1.2]{caverly_2019_lmi}.
A time-invariant system is a mapping from signals to signals, denoted
$\mbs{\mathcal{G}}: \ell_{2e} \to \ell_{2e}$, where $\ell_{2e}$ is the extended
inner product sequence space~\cite[\S B.1.1]{green_linear_1994}.
For LTI systems, $\mbs{\mathcal{G}}$ is a linear operator. A minimal state-space
realization of an LTI system is denoted~\cite[\S3.2.1]{green_linear_1994}
\begin{equation}
    \mbs{\mc{G}}
    \stackrel{\min}{\sim}
    \bma{c|c}
        \mbf{A} & \mbf{B} \\
        \hline 
        \mbf{C} & \mbf{D}
    \ema.
\end{equation}
Another way to realize an LTI system is through its transfer matrix,
\begin{equation}
    \mbf{y}(z) = \mbf{G}(z) \mbf{u}(z),
\end{equation}
where $\mbf{y}(z)$, $z \in \mathbb{C}$ denotes the $z$-transform of a
signal~\cite[\S 3.5.2]{antsaklis_2007_a}.
Evaluating the transfer matrix at
$z = e^{j \theta}$, where
$\theta = 2 \pi \Delta t f$
is the discrete-time frequency,
$\Delta t$ is the sampling period,
and $f$ is the continuous-time frequency, yields the frequency response of the
system.

\subsubsection{Norms of systems}
The size of a system can be quantified with a norm.
The $\mc{H}_\infty$ norm of a system $\mbs{\mc{G}}$ is the worst-case gain from
$\|\mbf{u}\|_2$ to ${\|\mbf{y}\|}_2 = {\|\mbs{\mc{G}}\mbf{u}\|}_2$.
That
is~\cite[\S B.1.1]{green_linear_1994},
\begin{equation}
    \|\mbs{\mc{G}}\|_\infty
    =
    \sup_{\mbf{u} \in \ell_{2}, \mbf{u} \neq \mbf{0}}
    \frac{\|\mbs{\mc{G}}\mbf{u}\|_2}{\|\mbf{u}\|_2}.%
    \label{eq:hinf-norm}
\end{equation}
In the frequency domain, this definition is equivalent to~\cite[\S
B.1.1]{green_linear_1994}
\begin{equation}
    \|\mbs{\mc{G}}\|_\infty
    =
    \sup_{\theta \in (-\pi, \pi]}
    \bar{\sigma}\left(\mbf{G}(e^{j \theta})\right),%
    \label{eq:hinf-norm-f}
\end{equation}
where $\bar{\sigma}(\cdot)$ denotes the maximum singular value of a matrix. The
\Hinf{}~norm of a system can be viewed as the peak magnitude of its frequency
response.

The \Htwo{}~norm is
\begin{equation}
    {\|\mbs{\mathcal{G}}\|}_2
    =
    \sqrt{%
        \frac{1}{2 \pi}
        \int_{-\pi}^{\pi}
        \trace{\left(
            {\mbf{G}(e^{j \theta})}^\herm
            \mbf{G}(e^{j \theta})
        \right)}
        \mathrm{d}\theta},
\end{equation}
where
$\mbf{G}(e^{j \theta})$ is the transfer matrix representation of
$\mbs{\mathcal{G}}$~\cite[\S4.4]{chen_1995_optimal}\cite{megretski_2004_interpretations}.
The \Htwo{}~norm can be viewed as the expected root-mean-squared (RMS) output of
a system when the input is unit variance white
noise~\cite[\S3.3.3]{green_linear_1994}\cite[\S5.7]{zhou_robust_1995}\cite{megretski_2004_interpretations}.

\subsubsection{Optimal controller synthesis}
The generalized plant, $\mbs{\mc{P}}$, depicted in
Figure~\ref{fig:generalized_plant_delta}, is central to the robust control
problem~\cite[\S3.8]{skogestad_2006_multivariable}. It consists of an
interconnection of the system to be controlled with a series of weighting
functions used to specify performance requirements.
The controller inputs and outputs, $\mbf{u}$ and $\mbf{y}$, represent the
signals accessible to a controller $\mbs{\mathcal{K}}$.
The performance inputs and outputs, $\mbf{w}_1$ and $\mbf{z}_1$, encode
performance requirements to be optimized.
Performance inputs could include disturbance and reference signals, while
performance outputs could include control effort and tracking error.
The uncertainty inputs and outputs, $\mbf{w}_2$ and $\mbf{z}_2$, connect the
generalized plant to the uncertainty block, $\mbs{\Delta}$, which can be any
system satisfying ${\|\mbs{\Delta}\|}_\infty \leq 1$.

In the absence of uncertainty, an optimal controller minimizes the impact of
$\mbf{w}_1$ on $\mbf{z}_1 = \mbs{\mc{T}}_{11} \mbf{w}_1$. That is, it solves
\begin{equation}
    \min_{\mbs{\mathcal{K}}}\;{\left\| \mbs{\mathcal{T}}_{11} \right\|}
\end{equation}
for some system norm of $\mbs{\mathcal{T}}_{11}$, like the \Hinf{}~norm or the
\Htwo{}~norm.
Weighting functions included inside the generalized plant allow performance
metrics, like disturbance rejection or tracking error, to be targeted over a
particular frequency band.
Posing an optimal control problem in this manner guarantees the asymptotic
stability of the closed-loop system as long as a system norm of
$\mbs{\mathcal{T}}_{11}$ is finite~\cite{zhou_robust_1995}.

Now consider the case where model uncertainty is present.
Let $\mbf{z}_2 = \mbs{\mc{T}}_{22} \mbf{w}_2$.
Assuming that $\mbs{\mathcal{K}}$ asymptotically stabilizes $\mbs{\mathcal{P}}$
for $\mbs{\Delta} = \mbf{0}$, the small-gain
theorem~\cite[\S4.9.4]{skogestad_2006_multivariable} implies that the
closed-loop system in Figure~\ref{fig:generalized_plant_delta} is asymptotically
stable for any ${\|\mbs{\Delta}\|}_\infty \leq 1$ if
${\|\mbs{\mathcal{T}}_{22}\|}_\infty < 1$.
This property is called \textit{robust stability}.
The mixed \Htwo{}-\Hinf{}~optimal controller achieves robust stability by
solving~\cite{kaminer_1993_mixed}.
\begin{align}
    \min_{\mbs{\mc{K}}}\;&{\left\| \mbs{\mathcal{T}}_{11} \right\|}_2
    \\
    \mathrm{s.t.}\;&{\left\| \mbs{\mathcal{T}}_{22} \right\|}_\infty < 1.
\end{align}
Let ${\begin{bmatrix} \mbf{z}_1^\trans& \mbf{z}_2^\trans \end{bmatrix}}^\trans =
    \mbs{\mathcal{T}} {\begin{bmatrix} \mbf{w}_1^\trans& \mbf{w}_2^\trans
    \end{bmatrix}}^\trans$.
The \Hinf{}~optimal controller also achieves robust stability by
solving~\cite[\S9.3]{skogestad_2006_multivariable}
\begin{equation}
    \gamma = \arg\min_{\mbs{\mc{K}}}\;{\left\| \mbs{\mathcal{T}} \right\|}_\infty,
\end{equation}
and verifying that $\gamma < 1$.

\subsubsection{Quantifying uncertainty}
To model the uncertainty for a given population of systems, an uncertainty
structure must be chosen and an uncertainty weighting function must be found.
Consider a nominal plant $\mbf{G}(z)$, a perturbed plant
$\mbf{G}_\mathrm{p}(z)$, and a perturbation, or residual, $\mbf{E}(z)$.
The perturbation can be expressed as~\cite[\S8.2.3]{skogestad_2006_multivariable}
\begin{equation}
    \mbf{E}(z) = \mbf{W}_2(z)\mbs{\Delta}(z)\mbf{W}_1(z),\quad
    {\|\mbs{\Delta}(z)\|}_\infty \leq 1,
\end{equation}
where often one of the weighting functions is set to identity.
These weights are placed inside the generalized plant, such that the input of
$\mbf{W}_1(z)$ is $\mbf{w}_2$ and the output of $\mbf{W}_2(z)$ is $\mbf{z}_2$.
The additive uncertainty model corresponds to~\cite[\S8.2.3]{skogestad_2006_multivariable}
\begin{equation}
    \mbf{G}_\mathrm{p}(z) = \mbf{G}(z) + \mbf{E}_\mathrm{a}(z).
\end{equation}
Let $\mbf{W}_2(z) = \mbf{1}$.
The remaining weighting function $\mbf{W}_1(z)$ is designed to
satisfy~\cite[\S8.2.3]{skogestad_2006_multivariable}
\begin{equation}
    {\|\mbf{G}_\mathrm{p}(z) - \mbf{G}(z)\|}_\infty
    \leq {\|\mbs{\Delta}(z)\|}_\infty\,{\|\mbf{W}_1(z)\|}_\infty
    \leq {\|\mbf{W}_1(z)\|}_\infty
\end{equation}
for all perturbed plants. Other common uncertainty forms
are~\cite[\S8.2.3]{skogestad_2006_multivariable}
\begin{itemize}
    \raggedright
    \item input multiplicative, {\small $\mbf{G}_\mathrm{p}(z) = \mbf{G}(z) \left(\mbf{1} + \mbf{E}_\mathrm{i}(z)\right)$},
    \item output multiplicative, {\small $\mbf{G}_\mathrm{p}(z) = \left(\mbf{1} + \mbf{E}_\mathrm{o}(z)\right) \mbf{G}(z)$},
    \item inverse additive, {\small $\mbf{G}_\mathrm{p}(z) = \mbf{G}(z){\left(\mbf{1} - \mbf{E}_\mathrm{ia}(z) \mbf{G}(z) \right)}^{-1}$},
    \item inverse input mult., {\small $\mbf{G}_\mathrm{p}(z) = \mbf{G}(z) {\left(\mbf{1} - \mbf{E}_\mathrm{ii}(z)\right)}^{-1}$}, and
    \item inverse output mult., {\small $\mbf{G}_\mathrm{p}(z) = {\left(\mbf{1} - \mbf{E}_\mathrm{io}(z)\right)}^{-1} \mbf{G}(z)$}.
\end{itemize}
Given a population of perturbed plants, each perturbation $\mbf{E}(z)$ is
typically calculated frequency-by-frequency. An uncertainty weight $\mbf{W}(z)$
that bounds all the perturbations is then designed and incorporated into the
generalized plant.

\subsection{Robust observer design}
Optimal observer design can be viewed as optimal controller design with a
particular choice of generalized plant. This section provides a brief overview
of linear observers and their relationship to linear controllers.
Consider the state-space representation of a strictly proper LTI system,
\begin{align}
    \mbf{x}_{k + 1}
    &=
    \mbf{A}
    \mbf{x}_k
    +
    \mbf{B}
    \mbf{u}_k,
    \\
    \mbf{y}_k
    &=
    \mbf{C}
    \mbf{x}_k,
\end{align}
where $(\mbf{A},\mbf{C})$ is observable.
The structure of the most common type of observer, the \textit{Luenberger
    observer}, is~\cite{marquez_2005_robust_state_observer}
\begin{align}
    \mbfhat{x}_{k+1}
    &=
    \mbf{A}
    \mbfhat{x}_k
    +
    \mbf{B}
    \mbf{u}_k
    +
    \mbf{L}
    \left(\mbf{y}_k - \mbfhat{y}_k\right),
    \\
    \mbfhat{y}_k &= \mbf{C} \mbfhat{x}_k,
\end{align}
where $\mbf{L}$ is the observer gain.
The observer gain is chosen to asymptotically stabilize the error
dynamics~\cite{marquez_2005_robust_state_observer}
\begin{equation}
    \mbf{e}_{k + 1} = \left(\mbf{A} - \mbf{L}\mbf{C}\right) \mbf{e}_k,
\end{equation}
where $\mbf{e}_k = \mbfhat{x}_k - \mbf{x}_k$.

\begin{figure}[htbp]
    \centering
    \scalebox{0.75}{\begin{tikzpicture}
        \node[
            block_ob,
        ] (K) at (0, 0) {$\mbs{\mc{K}}$};
        \node[
            block_ob,
            below left = 1.8cm and 1.2cm of K,
        ] (Gp) {$\mbs{\mc{G}}$};
        \node[
            block_ob,
            below right = 1.8cm and 1.2cm of K,
        ] (Cp) {$\mbs{\mc{C}}$};

        \node[
            diff_ns,
            right = 2.4cm of K,
        ] (s2) {};
        \node[
            sum_ne,
            left = 2.4cm of K,
        ] (s4) {};
        \draw[bdarrow] (K) -- (s4)
            node[pos=0.2, above] {$\mbf{y}_k^{\mathrm{c}}$};
        \draw[bdarrow] (s4) |- (Gp);
        \draw[bdarrow] (Gp) -- (Cp);
        \draw[bdarrow] (s2) -- (K);
        \draw[bdarrow] (Cp) -| (s2)
            node[pos=0.8, right] {$\mbfhat{y}_k$};
        \draw[bdarrow] (s2) ++(0.0cm, 1.2cm) -- (s2)
            node[pos=0, right] {$\mbf{y}_k$};
        \draw[bdarrow] (s4) ++(0.0cm, 1.2cm) -- (s4)
            node[pos=0, right] {$\mbf{u}_k$};
        \draw[bdarrow] (Gp) ++(1.2cm, 0cm) |-  ++(0.6cm, -1.2cm) (Gp)
            node[pos=1, right] {$\hat{\mbf{x}}_k$};
    \end{tikzpicture}}
    \caption{%
    The structure of an input-output observer. Note that the output
    matrix $\mbf{C}$ is drawn as a separate system from $\mbs{\mathcal{G}}$ to
    allow access to the state estimate $\mbfhat{x}_k$.}\label{fig:obs_structure}
\end{figure}
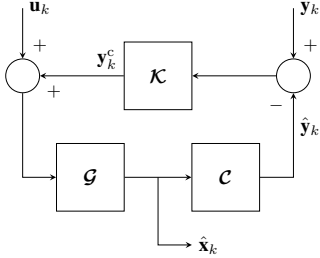
However, observers need not be limited to static gains.
In~\cite{marquez_2003_a_frequency_domain,marquez_2005_robust_state_observer},
robust optimal control techniques are used to synthesize \textit{input-output
    observers} of the form
\begin{align}
    \mbf{x}_{k + 1}^\mathrm{c}
    &=
    \mbf{A}^{\!\mathrm{c}}
    \mbf{x}_k^\mathrm{c}
    +
    \mbf{B}^\mathrm{c}
    \left(\mbf{y}_k - \mbfhat{y}_k\right),
    \\
    \mbf{y}_k^\mathrm{c}
    &=
    \mbf{C}^\mathrm{c}
    \mbf{x}_k^\mathrm{c}
    +
    \mbf{D}^\mathrm{c}
    \left(\mbf{y}_k - \mbfhat{y}_k\right),
    \\
    \mbfhat{x}_{k + 1}
    &=
    \mbf{A}
    \mbfhat{x}_k
    +
    \mbf{B}
    \left(\mbf{u}_k + \mbf{y}_k^\mathrm{c}\right),
    \\
    \mbfhat{y}_k
    &=
    \mbf{C}
    \mbfhat{x}_k.
\end{align}
A block diagram of an input-output observer is shown in
Figure~\ref{fig:obs_structure}, where
\begin{equation}
    \mbs{\mc{K}}
    \stackrel{\min}{\sim}
    \bma{c|c}
        \mbf{A}^{\!\mathrm{c}} & \mbf{B}^\mathrm{c} \\
        \hline 
        \mbf{C}^\mathrm{c} & \mbf{D}^\mathrm{c}
    \ema,
    \quad
    \mbs{\mc{C}}
    \mbs{\mc{G}}
    \stackrel{\min}{\sim}
    \bma{c|c}
        \mbf{A} & \mbf{B} \\
        \hline 
        \mbf{C} & \mbf{0}
    \ema.\label{eq:def_CG}
\end{equation}
Inside the observer, a copy of the nominal plant is used for comparison against
measurements. The synthesized controller $\mbs{\mc{K}}$ compares the
measurements with the nominal plant's output and makes corrections to its
input accordingly.
Note that the output matrix $\mbf{C}$ is drawn as a separate system
$\mbs{\mathcal{C}}$ in Figure~\ref{fig:obs_structure}, to allow access to the
state estimate $\mbfhat{x}_k$ in the block diagram.
An example of a generalized plant for an input-output observer design problem
can be found later in Figure~\ref{fig:obs_generalized_plant}.

Unlike a controller, an observer cannot destabilize the observed system. As
such, robustness in an observer does not refer to robust stability.
Instead, as in~\cite{marquez_2003_a_frequency_domain,
marquez_2005_robust_state_observer}, robustness in an observer refers to
insensitivity of performance to model uncertainty.
Loosely, the uncertainty weight is therefore used to inform the observer of when
it can trust the predictions of its internal nominal model. When uncertainty is
high, the nominal model may not match the true system, so external measurements
should be used to correct the model predictions. When uncertainty is low, the
nominal model is likely to match the true systems, so noise and disturbances can
be rejected by using the observer's internal model without correction.

\subsection{Koopman operator theory}
This section provides a brief overview of Koopman operator theory with exogenous
inputs.

\subsubsection{The Koopman operator}
Consider the nonlinear difference equation
\begin{equation}
    \mbf{x}_{k+1} = \mbf{f}(\mbf{x}_{k}, \mbf{u}_k),\label{eq:dynamics_input}
\end{equation}
where the state is
$\mbf{x}_{k} \in \mc{M} \subseteq \mathbb{R}^{m \times 1}$
and the input is
$\mbf{u}_{k} \in \mc{N} \subseteq \mathbb{R}^{n \times 1}$.
Also consider the set $\mc{H}$ of all scalar-valued \textit{lifting functions},
$\psi: \mc{M} \times \mc{N} \to \mathbb{R}$. The \textit{Koopman operator},
$\mc{U}: \mc{H} \to \mc{H}$,
composes lifting functions with $\mbf{f}(\cdot)$, advancing them in time by one
timestep. That is,
\begin{equation}
    (\mc{U} \psi)(\mbf{x}_{k}, \mbf{u}_{k})
    = \psi(\mbf{f}(\mbf{x}_{k}, \mbf{u}_k), \star),
    \label{eq:koopman_def_input}
\end{equation}
where
$\star = \mbf{u}_k$
if the input has state-dependent dynamics, or
$\star = \mbf{0}$
if the input has no dynamics~\cite[\S6.5]{kutz_dynamic_2016}.
The set $\mc{H}$ is often an infinite-dimensional Hilbert space. Sometimes this
space is explicitly chosen, while other times, it is implicitly defined by the
lifting functions~\cite[\S2.1]{brunton_2022_modern}.

To approximate the Koopman operator in finite dimensions, a finite subset of
lifting functions must be chosen.
Let the \textit{vector-valued lifting functions}
$\mbs{\psi}: \mc{M} \times \mc{N} \to \mathbb{R}^{p \times 1}$
be written as
\begin{equation}
    \mbs{\psi}(\mbf{x}_k, \mbf{u}_k) = \begin{bmatrix}
        \mbs{\vartheta}(\mbf{x}_k) \\
        \mbs{\upsilon}(\mbf{x}_k, \mbf{u}_k)
    \end{bmatrix},
\end{equation}
where the state-dependent lifting functions are
$\mbs{\vartheta}: \mc{M} \to \mathbb{R}^{p_\vartheta \times 1}$,
the input-dependent lifting functions are
$\mbs{\upsilon}: \mc{M} \times \mc{N} \to \mathbb{R}^{p_\upsilon \times 1}$,
and
$p_\vartheta + p_\upsilon = p$.
A finite-dimensional approximation of~\eqref{eq:koopman_def_input}
is~\cite[\S6.5.1]{kutz_dynamic_2016}
\begin{equation}
    \mbs{\vartheta}(\mbf{x}_{k+1}) \\
    =
    \mbf{U}
    \mbs{\psi}(\mbf{x}_{k}, \mbf{u}_{k})
    + \mbs{\epsilon}_k,%
    \label{eq:U_part}
\end{equation}
where $\mbf{U} = \begin{bmatrix} \mbf{A} & \mbf{B} \end{bmatrix}$ is the
\textit{Koopman matrix}.
Expanding~\eqref{eq:U_part} results in the linear state-space form,
\begin{equation}
    \mbs{\vartheta}(\mbf{x}_{k+1})
    =
    \mbf{A} \mbs{\vartheta}(\mbf{x}_{k})
    + \mbf{B} \mbs{\upsilon}(\mbf{x}_k, \mbf{u}_k)
    + \mbs{\epsilon}_k.\label{eq:koopman_state_space}
\end{equation}

\subsubsection{Data-driven Koopman operator approximation}
To approximate the Koopman matrix from a dataset
$\mc{D} = {\{\mbf{x}_k, \mbf{u}_k\}}_{k=0}^q$, lifting functions must first be
chosen.
Koopman lifting functions can be inspired by the dynamics of the system in
question~\cite{abraham_active_2019, mamakoukas_local_2019, kaiser_data_2021},
taken from a standard set of basis functions like polynomials, sinusoids, or
radial basis functions~\cite{bruder_modeling_2019, abraham_model-based_2017,
    mallen_koopman_2023, susuki_2024_control, svec_2023_predictive}, or chosen to
approximate a given kernel~\cite{guo_2022_koopman, degennaro_2019_scalable,
    rahimi_2007_random}. Time-delayed states are also often included in the lifted
state~\cite{bruder_modeling_2019, korda_2018_linear, pan_2020_structure}.

For a given choice of lifting functions, consider the lifted snapshot matrices
\begin{align}
    \mbs{\Psi} &= \begin{bmatrix}
        \mbs{\psi}_{0} & \mbs{\psi}_{1} & \cdots & \mbs{\psi}_{q-1}
    \end{bmatrix} \in \mathbb{R}^{p \times q}, \\
    \mbs{\Theta}_+ &= \begin{bmatrix}
        \mbs{\vartheta}_{1} & \mbs{\vartheta}_{2} & \cdots & \mbs{\vartheta}_{q}
    \end{bmatrix} \in \mathbb{R}^{p_\vartheta \times q},\label{eq:Theta}
\end{align}
where
$\mbs{\psi}_k = \mbs{\psi}(\mbf{x}_k, \mbf{u}_k)$
and
$\mbs{\vartheta}_k = \mbs{\vartheta}(\mbf{x}_k)$.

Least-squares is the simplest way to approximate the Koopman matrix from data.
Minimizing $\mbs{\epsilon}_k$ leads to the optimization problem
\begin{equation}
    \min_{\mbf{U}}\;\frac{1}{q} {\left\|\mbs{\Theta}_+ - \mbf{U}
    \mbs{\Psi}\right\|}_\frob^2\label{eq:koopman-cost},
\end{equation}
whose solution is~\cite[\S1.2.1]{kutz_dynamic_2016}
\begin{equation}
    \mbf{U} = \mbs{\Theta}_+ \mbs{\Psi}^\dagger.\label{eq:pseudoinverse}
\end{equation}

\textit{Extended dynamic mode decomposition} (EDMD)~\cite{williams_data-driven_2015}
improves performance and numerical conditioning in the least-squares problem
when the dataset contains many fewer states than snapshots (\textit{i.e.}, when
$p \ll q$)~\cite[\S10.3]{kutz_dynamic_2016}. Assuming $\mbs{\Psi}$ is full rank,
the EDMD approximation of the Koopman matrix
is
\begin{equation}
    \mbf{U}
    =
    \mbs{\Theta}_+
    \left(
        \mbs{\Psi}^\trans
        \mbs{\Psi}^{\trans^\dagger}
    \right)
    \mbs{\Psi}^\dagger
    =
    \left(\mbs{\Theta}_+ \mbs{\Psi}^\trans\right)
    {\left(\mbs{\Psi} \mbs{\Psi}^\trans\right)}^\dagger
    =
    \mbf{G}
    \mbf{H}^\dagger,
    \label{eq:edmdi-sol}
\end{equation}
where
\begin{equation}
    \mbf{G}
    =
    \frac{1}{q}
    \mbs{\Theta}_+ \mbs{\Psi}^\trans \in \mathbb{R}^{p_\vartheta \times p},
    \quad
    \mbf{H}
    =
    \frac{1}{q}
    \mbs{\Psi} \mbs{\Psi}^\trans \in \mathbb{R}^{p \times p}.\label{eq:edmdi-sol-scaled-H}
\end{equation}
Note that, when the columns of $\mbs{\Psi}$ are linearly independent, $\mbf{H} =
    \mbf{H}^\trans > 0$.

\section{Methodology}\label{sec:methodology}
The proposed robust nonlinear observer synthesis method applies the input-output
observer structure
of~\cite{marquez_2003_a_frequency_domain,marquez_2005_robust_state_observer} to
a population of Koopman models. The steps of the proposed method are as follows.
\begin{enumerate}
    \item Identify a Koopman model for every plant in a population using the
          same set of lifting functions.
    \item Select a nominal plant and an uncertainty form, compute residuals, and
          design an uncertainty weighting function to bound the residuals in the
          frequency domain.
    \item Design performance weighting functions, form the generalized plant,
          and synthesize a robust controller (\textit{e.g.}, \Hinf{}, mixed
          \Htwo{}-\Hinf).
    \item Interconnect the controller with the nominal Koopman plant as in
          Figure~\ref{fig:obs_structure}.
\end{enumerate}

Thanks to the linearity of the Koopman operator, the input-output observer
structure requires little modification, as the observer synthesis process takes
place in the lifted space. This is viewed as a positive aspect of the approach,
as existing robust control and estimation techniques are powerful and
well-studied.

One key difference between the proposed Koopman observer and a linear observer
is how the nominal model is used when the observer is deployed.
During synthesis, the Koopman model is treated an LTI system.
However, during operation, the original system's state is retracted and
re-lifted at each timestep.
This prevents the system interconnection in Figure~\ref{fig:obs_structure} from
being collapsed into a single LTI block, but yields more accurate model
predictions, as it ensures that the structure of the lifted state is always
respected.
This approach results in the \textit{local error} definition
of~\cite{mamakoukas_2020_learning}.
Note that when comparing Koopman observers to their linear counterparts,
weights on lifted states should be zero to ensure that both methods minimize the
same cost functions.

\section{Experimental Validation}\label{sec:experimental}
\begin{figure}[htbp]
    \centering
    \includegraphics[width=2.4in]{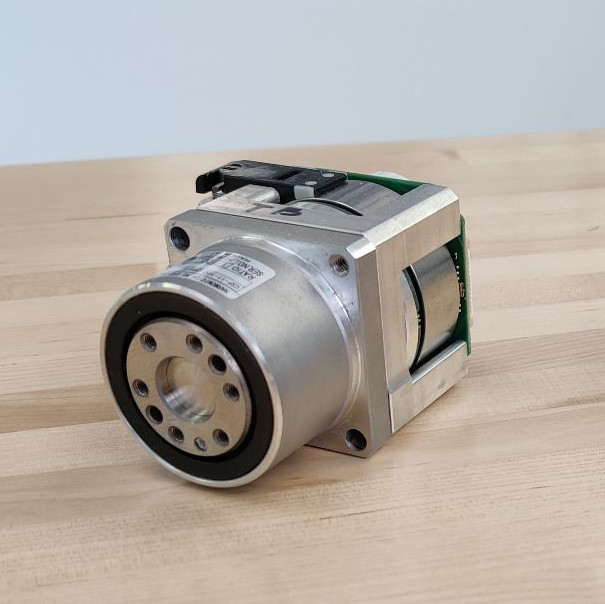}
    \caption{%
    Motor drive used to generate the training data, which consists of
    a motor with a Harmonic Drive gearbox. The gearbox introduces nonlinear
    oscillations into the system, leading to tracking errors at specific
    frequencies related to the input velocity. Photo courtesy of Alexandre
    Coulombe.}\label{fig:drive}
\end{figure}

In this section, the proposed robust nonlinear observer synthesis methodology is
applied to a population of 38 motor drives.
Each motor drive, pictured in Figure~\ref{fig:drive}, consists of an electric
motor and a Harmonic Drive gearbox. The gearbox introduces a periodic
oscillation into the system which cannot be captured by a linear
model~\cite{ghorbel_1998_kinematic, tuttle_1996_nonlinear}.
A Koopman model for each drive is identified, manufacturing uncertainty is
characterized in the frequency domain, and a robust Koopman observer is designed
to estimate motor velocity and current using position measurements.
Each motor drive is identified with and without an asymmetric inertial load. The
models used for observer synthesis are identified using unloaded drives, so the
motor current estimation errors can be treated as load torque estimates.
The uncertainty characterization used for robust observer design can also be
used to detect incorrectly installed motor drives. An outlier detection approach
based on the uncertainty characterization done in this section is discussed in
Appendix~\ref{sec:app_outlier}.

The full motor drive dataset is available for download
at~\cite{dahdah_2024_quantifying}. A datasheet following the format
of~\cite{gebru_2021_datasheets} is also provided.
The software required to fully reproduce the results of this paper is available
at \url{https://github.com/decargroup/robust_observer_koopman}. This code
extends \texttt{pykoop}~\cite{pykoop}, the authors' open-source Koopman operator
approximation library for Python.

\subsection{Dataset}
The motor drives under consideration operate in closed-loop, accepting position
and velocity reference signals and returning position, velocity, and current
measurements. The velocity measurements are computed from position measurements
by the drives via a filtered finite difference scheme.
The dataset consists of 40 episodes per drive, each approximately
\SI{20}{\second} long. The drive is loaded with an asymmetric inertial load for
half of the episodes.
The loaded and unloaded datasets are split into 18 training episodes and two
test episodes.
The reference position, reference velocity, measured position, measured
velocity, and measured motor current are recorded at \SI{1}{\kilo\hertz}.
The positions and velocities are recorded in \si{\radian} and
\si{\radian/\second} respectively at the output shaft of the gearbox, while the
current is recorded as a fraction of the drive's full-scale current. The
gearboxes under consideration reduce input velocities by a factor of 100.

The drive's control software accepts position checkpoints and generates the
smoothed trapezoidal velocity trajectories required to reach each checkpoint.
Each episode contains 10 pseudorandom position checkpoints set within one
revolution of the gearbox output shaft in either direction. The resulting
velocity profile resembles a smoothed pseudorandom binary sequence
(PRBS)~\cite[\S13.3]{ljung_1999_system}. For all episodes, the maximum allowed
velocity and acceleration settings are set in the drive's control software to
generate the most challenging training and test trajectories possible.

\subsection{Koopman operator approximation}
Koopman operator identification for a single motor drive is now considered.
Lifting functions for the motor drive are chosen based on the nonlinear
oscillations that are known to be present in Harmonic Drive
gearboxes~\cite{ghorbel_1998_kinematic, tuttle_1996_nonlinear}.
This oscillation is kinematic in nature~\cite{ghorbel_1998_kinematic,
tuttle_1996_nonlinear} with a fixed phase, which can be treated as a calibration
parameter that must be identified for each motor drive.
EDMD is then used to identify the Koopman matrix given the calibrated lifting
functions.

\subsubsection{Lifting function selection}
Harmonic Drive gearboxes are known to generate vibrations at a frequency once
and twice the input frequency~\cite{tuttle_1996_nonlinear,
    ghorbel_1998_kinematic}.
These oscillations are first characterized experimentally, and then used to
inform the design of Koopman lifting functions.

Let $\theta(t)$ be the gearbox output angle in \si{\radian} and let $i(t)$ be
the measured motor current as a fraction of the full-scale current.
The nonlinear vibration can be modelled as an exogenous load torque disturbance,
which is proportional to the current disturbance~\cite{tuttle_1996_nonlinear}
\begin{equation}
    i{^\mathrm{d}}(t)
    =
    a_1 \sin{(r\theta(t) + \varphi_1)}
    +
    a_2 \sin{(2r\theta(t) + \varphi_2)},\label{eq:hd_exogenous}
\end{equation}
where $a_1$ and $a_2$ are vibration amplitudes, $\varphi_1$ and
$\varphi_2$ are phase shifts, and $r$ is the reduction ratio of the gearbox.
\begin{figure}[htbp]
    \centering
    \includegraphics[width=3.2in]{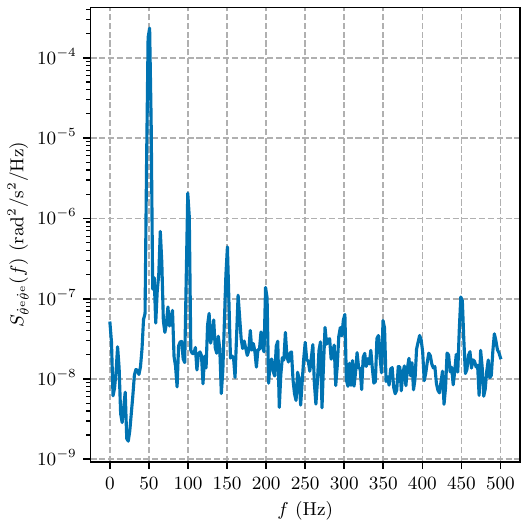}
    \caption{%
    Power spectral density of output velocity tracking error during a
    constant-velocity trajectory segment. The velocity at the gearbox input is
    \SI{50}{rev\per\second}, leading to vibrations at integer multiples of
    \SI{50}{\hertz}. The most prominent tracking errors occur at the
    fundamental frequency. A logarithmic scale is used to better show the
    high-frequency harmonics.}\label{fig:error_fft}
\end{figure}
These vibrations appear as tracking errors in the position and velocity of the
drive. They also appear in the current command due to the feedback action of the
controller.
Figure~\ref{fig:error_fft} shows the power spectral density of a drive's
velocity tracking error during a constant \SI{0.5}{rev/s} movement, which is the
drive's maximum speed. Since the gearbox reduction ratio is known to be $r=100$,
the resulting gearbox input velocity is \SI{50}{rev/s}. Velocity tracking errors
occur at integer multiples of \SI{50}{\hertz}, with decreasing power as
frequency increases. By several orders of magnitude, the most significant
tracking errors occur at \SI{50}{\hertz}.

Inspired by Figure~\ref{fig:error_fft} and by~\eqref{eq:hd_exogenous}, the
Koopman lifting functions for the motor drive system are chosen to be
\begin{align}
    \mbs{\vartheta}_k
    &=
    \begin{bmatrix}
        \theta_k \\
        \dot{\theta}_k \\
        \sin{(100\,\theta_k + \varphi)}
    \end{bmatrix},\label{eq:state_dependent_lifting}
    \\
    \mbs{\upsilon}_k
    &=
    i_k.\label{eq:input_dependent_lifting}
\end{align}
Since the \SI{50}{\hertz} error in Figure~\ref{fig:error_fft} is by far the
largest, only that sinusoidal term is included in the lifting functions.
It was found experimentally that including higher-frequency harmonics has a
minimal effect on the system's overall prediction error.
The vibration amplitude $a_1$ is identified as part of the approximated
Koopman matrix.
Note the inclusion of a constant phase offset $\varphi$
in~\eqref{eq:state_dependent_lifting}, which must be computed for each drive.
A simple calibration procedure for $\varphi$ is discussed in
Appendix~\ref{sec:app_calibration}.

\begin{figure}[htbp]
    \centering
    \includegraphics[width=3.2in]{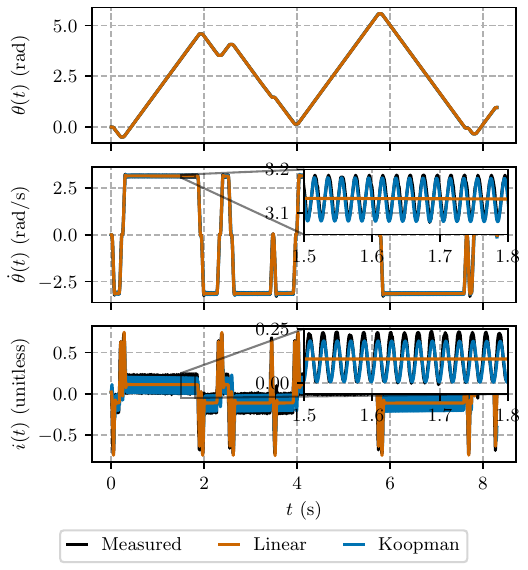}
    \caption{%
    Predicted position, velocity, and current trajectories for linear
    and Koopman drive models using the first test episode. The linear model is
    not able to reproduce the Harmonic Drive oscillations, and instead predicts
    the average velocity and current. The Koopman model is able to accurately
    predict the oscillations.}\label{fig:model_predictions_traj}
\end{figure}
\begin{figure}[htbp]
    \centering
    \includegraphics[width=3.2in]{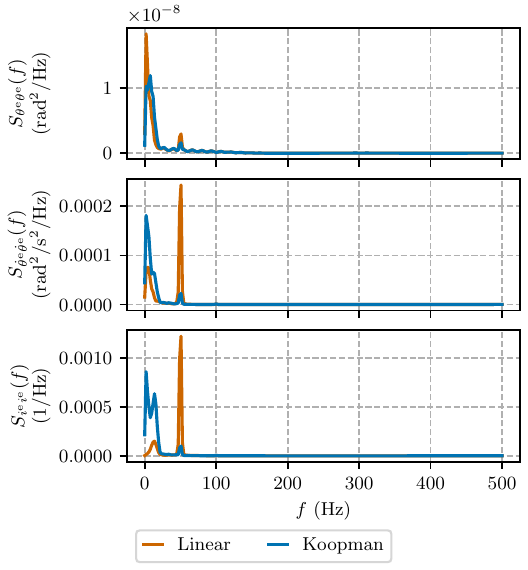}
    \caption{Predicted position, velocity, and current error power spectral
    densities for linear and Koopman drive models using the first test episode.
    The Koopman model better predicts the gearbox oscillations, leading to lower
    velocity and current errors at \SI{50}{\hertz}. A linear scale is used to
    emphasize the difference in error.}\label{fig:model_predictions_fft}
\end{figure}

\begin{figure*}[htbp]
    \centering
    \begin{subfigure}[t]{3.2in}
        \centering
        \includegraphics[width=3.2in]{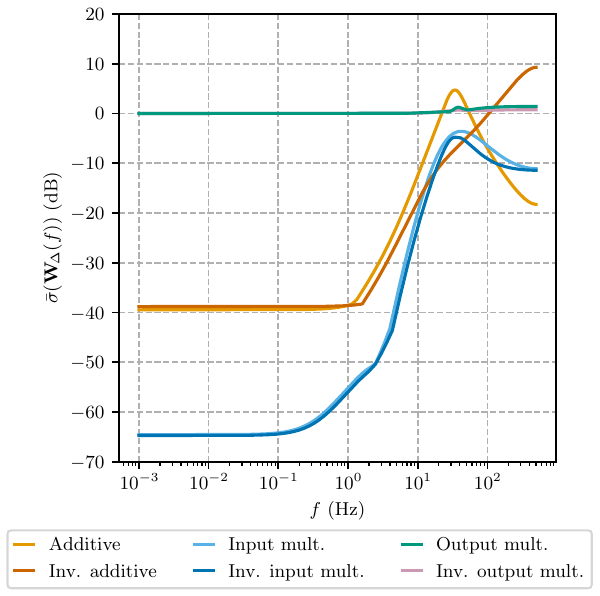}
        \caption{Linear model.}\label{fig:uncertainty_linear}
    \end{subfigure}%
    \hfill
    \begin{subfigure}[t]{3.2in}
        \centering
        \includegraphics[width=3.2in]{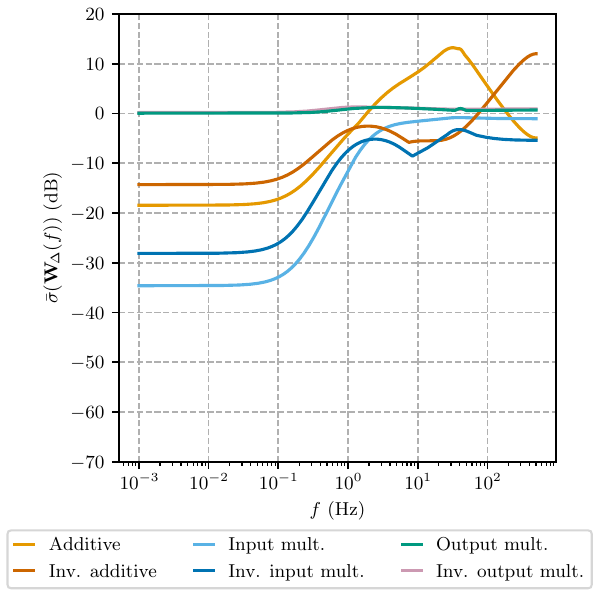}
        \caption{Koopman model.}\label{fig:uncertainty_koopman}
    \end{subfigure}
    \caption{%
    Upper bounds on the residuals for each uncertainty form. Both
    output multiplicative uncertainty forms have high uncertainty everywhere,
    while both input multiplicative forms have uncertainty below
    \SI{0}{\deci\bel} everywhere. Inverse input multiplicative uncertainty is
    preferred, as it remains lower than its feedforward counterpart at higher
    frequencies.}\label{fig:uncertainty}
\end{figure*}
\begin{figure*}[htbp]
    \centering
    \begin{subfigure}[t]{3.2in}
        \centering
        \includegraphics[width=3.2in]{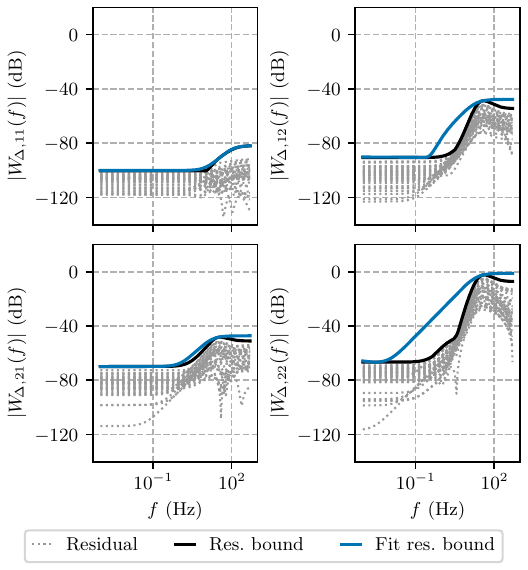}
        \caption{Linear uncertainty bounds,
        input-to-output.}\label{fig:uncertainty_bound_mimo_linear}
    \end{subfigure}%
    \hfill
    \begin{subfigure}[t]{3.2in}
        \centering
        \includegraphics[width=3.2in]{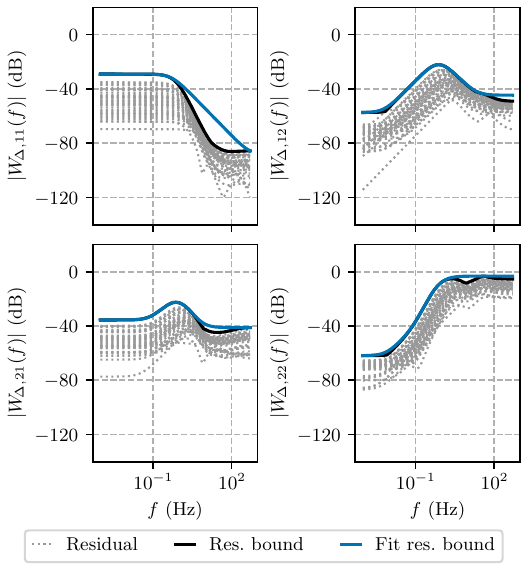}
        \caption{Koopman uncertainty bounds,
        input-to-output.}\label{fig:uncertainty_bound_mimo_koopman}
    \end{subfigure}
    \caption{Inverse input multiplicative uncertainty bounds and fit transfer
    functions for linear and Koopman models.}\label{fig:uncertainty_bound}
\end{figure*}

\subsubsection{Koopman regression}
EDMD with Tikhonov regularization~\cite{tikhonov_1995_numerical} is used to
identify Koopman matrices using linear lifting functions and the lifting
functions~\eqref{eq:state_dependent_lifting}
and~\eqref{eq:input_dependent_lifting}.
A manual bisection procedure is used to find the smallest regularization
coefficient that results in stable Koopman systems for each drive, both with
linear and nonlinear lifting functions.
The Tikhonov regularization coefficient required to stabilize all Koopman models
is $\alpha=90$.
More advanced methods like those in~\cite{dahdah_system_2022} could also be
used to guarantee asymptotic stability without such a high regularization
coefficient. However, in spite of the large regularization coefficient, the
identified models are shown to result in good predictive performance and
observer designs.

\subsubsection{Prediction results}
The predicted position, velocity, and current trajectories of one motor drive
are shown in Figure~\ref{fig:model_predictions_traj}. The corresponding
prediction errors are shown in Figure~\ref{fig:model_predictions_fft} in the
frequency domain.
Figure~\ref{fig:model_predictions_traj} shows that the linear model correctly
identifies the low-frequency dynamics of the motor drive, but does not predict
the vibrations induced by the Harmonic Drive gearbox. In contrast, the Koopman
model predicts the oscillations accurately.
Figure~\ref{fig:model_predictions_fft} shows the power spectral density of the
prediction errors, which demonstrate that the Koopman model significantly
reduces prediction errors at \SI{50}{\hertz}, while introducing more
low-frequency current prediction error.
While higher accuracy could be obtained by using a more complex Koopman model,
the simplicity of the proposed model is advantageous because it will lead to a
lower-order observer design.

\subsection{Uncertainty characterization}
Now that a Koopman model for each motor drive has been identified, uncertainty
is quantified within the population.
Rather than computing an average model to use as the nominal plant, one motor
drive model from the population is chosen as the nominal plant.
For each uncertainty form and for each choice of nominal plant, residual
transfer matrices are computed. The uncertainty form and nominal plant yielding
the lowest peak residual is selected for observer design.
%
%
%
For the sake of comparison, the same nominal plant is used for the linear and
Koopman approaches, selected using the Koopman model's residuals.

Figure~\ref{fig:uncertainty} shows the upper bounds on the maximum singular
values of the residuals for each uncertainty form. For each uncertainty form,
the nominal plant yielding the lowest uncertainty is shown.
Both output multiplicative uncertainty forms
have high gain over the whole frequency spectrum, and are therefore unsuitable
for controller or observer design. The input multiplicative uncertainty forms
are more desirable, as they have low uncertainty at low frequency and their
gains remain below \SI{0}{\deci\bel} over all frequencies. Between the two, the
inverse input multiplicative uncertainty form has the lowest gain at high
frequencies, so it is selected for observer design.

The linear inverse input multiplicative uncertainty bound in
Figure~\ref{fig:uncertainty} has lower uncertainty than the Koopman uncertainty
bound at low frequencies, but similar uncertainty at high frequencies. Both
uncertainty bounds have sufficiently low uncertainty for controller or observer
design at low frequencies.
%

The uncertainty bounds in Figure~\ref{fig:uncertainty} are computed
frequency-by-frequency.
Once an uncertainty model is selected, transfer functions must be designed to
bound them. A nonlinear optimization problem is solved to find the transfer
function coefficients that result in a magnitude response that closely bounds
the residuals at each frequency.
Figure~\ref{fig:uncertainty_bound} shows the uncertainty weights bounding the
linear and Koopman residuals. Individual weighting functions are computed for
each input-to-output transfer function. For the linear model, first- and
second-order transfer functions are used. The Koopman models require third-order
transfer functions, except for $W_{\Delta,11}(z)$, which uses a first-order
transfer function.
In both the linear and Koopman cases, $W_{\Delta,22}(z)$ represents the majority of
the uncertainty in the transfer matrix.
While the linear and Koopman uncertainty bounds have similar magnitudes at high
frequencies, the Koopman uncertainty bound is easier to fit closely with a
low-order transfer function.

\subsection{Robust observer design}
Robust observers are now synthesized for both linear and Koopman models of the
motor drive population.
Since the motor drive has no true velocity sensor, motor velocity and current
are observed given position measurements. Because the observer is synthesized
using unloaded models, the motor current can be viewed as a proxy for the
motor's electromagnetic torque. When the observer is fed measurements from a
loaded drive, it predicts only the electromagnetic torque, not the load torque.
Consequently, the current prediction error can be viewed as an estimate of the
motor drive's load torque.

\subsubsection{Observer generalized plant}
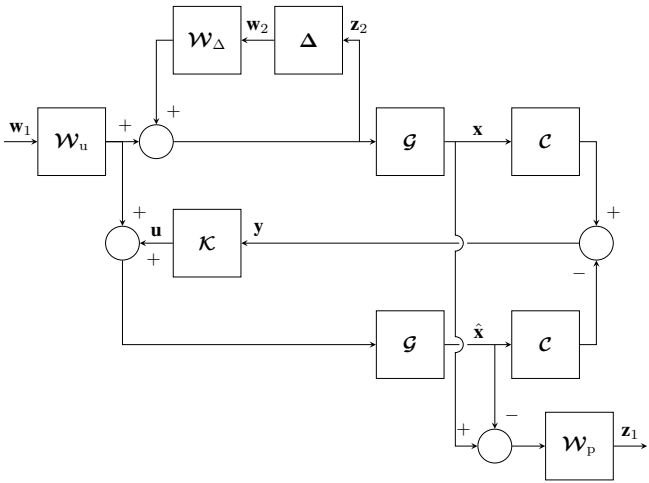
\begin{figure}[htbp]
    \centering
    \scalebox{0.75}{\begin{tikzpicture}
        \node[
            block_ob,
        ] (G) at (0, 0) {$\mbs{\mc{G}}$};
        \node[
            block_ob,
            right = 2.4cm of G,
        ] (C) {$\mbs{\mc{C}}$};
        \node[
            block_ob,
            above left = 1.8cm and 1.8cm of G,
        ] (Delta) {$\mbs{\Delta}$};
        \node[
            block_ob,
            left = 1.8cm of Delta,
        ] (W_Delta) {$\mbs{\mc{W}}_\Delta$};
        \node[
            block_ob,
            left = 6.0cm of G,
        ] (Wu) {$\mbs{\mc{W}}_\mathrm{u}$};
        \node[
            block_ob,
            below left = 1.8cm and 3.6cm of G,
        ] (K) {$\mbs{\mc{K}}$};
        \node[
            block_ob,
            below = 3.6cm of G,
        ] (Gp) {$\mbs{\mc{G}}$};
        \node[
            block_ob,
            below = 3.6cm of C,
        ] (Cp) {$\mbs{\mc{C}}$};
        \node[
            sum_nw,
            right = 1.5cm of Wu,
        ] (s1) {};
        \node[
            sum_ne,
            below left = 1.8cm and 0.6cm of s1,
        ] (s4) {};
        \node[
            diff_ns,
            right = 8.4cm of s4,
        ] (s2) {};
        \node[
            diff_nw,
            below left = 1.8cm and 0.9cm of Cp,
        ] (s5) {};
        \node[
            block_ob,
            right = 1.5cm of s5,
        ] (Wp) {$\mbs{\mc{W}}_\mathrm{p}$};
        \draw[bdarrow] (Wu) -- (s1);
        \draw[bdarrow] (Wu) -| (s4);
        \draw[bdarrow] (s1) -- (G);
        \draw[bdarrow] (K) -- (s4)
            node[pos=0.5, above] {$\mbf{u}$};
        \draw[bdarrow] (s4) |- (Gp);
        \draw[bdarrow] (C) -| (s2);
        \draw[bdarrow] (G) -- (C)
            node[midway, above] {$\mbf{x}$};
        \draw[bdarrow] (W_Delta) -| (s1);
        \draw[bdarrow] (Delta) -- (W_Delta)
            node[midway, above] {$\mbf{w}_2$};
        \draw[bdarrow] (s1) ++(3.6cm, 0.0cm) |- (Delta)
            node[midway, above] {$\mbf{z}_2$};
        \draw[bdarrow] (Wu) ++(-1.2cm, 0.0cm) -- (Wu)
            node[midway, above] {$\mbf{w}_1$};
        \draw[bdarrow] (s5) -- (Wp);
        \draw[bdarrow] (Wp) -- ++(1.2cm, 0.0cm) (Wp)
            node[midway, above] {$\mbf{z}_1$};
        %
        \draw[bdarrow] (Cp) -| (s2);
        \draw[bdarrow] (Cp) ++(-0.9cm, 0.0cm) -- (s5);
        \path[spath/save=a] (s2) -- (K);
        \path[spath/save=b] (G) ++(0.8cm, 0cm) |- (s5);  
        \path[spath/save=c] (Gp) -- (Cp);
        \path[spath/save=arc1] (0,0) arc[radius=1cm, start angle=180, delta angle=-180];
        \path[spath/save=arc2] (0,0) arc[radius=1cm, start angle=180, delta angle=-180];
        \tikzset{
          spath/split at intersections with={b}{a},
          spath/insert gaps after components={b}{8pt},
          spath/join components with={b}{arc1},
          spath/split at intersections with={a}{b},
          spath/insert gaps after components={a}{4pt},
        }
        \tikzset{
          spath/split at intersections with={b}{c},
          spath/insert gaps after components={b}{8pt},
          spath/join components with={b}{arc2},
          spath/split at intersections with={c}{b},
          spath/insert gaps after components={c}{4pt},
        }
        \draw[bdarrow,spath/use=a]
            node[pos=0.96, above] {$\mbf{y}$};
        \draw[bdarrow,spath/use=b];
        \draw[bdarrow,spath/use=c]
            node[pos=0.63, above] {$\hat{\mbf{x}}$};
    \end{tikzpicture}}
    \caption{The generalized plant for a robust input-output observer problem with
    inverse input multiplicative uncertainty.}\label{fig:obs_generalized_plant}
\end{figure}

The generalized plant used for robust observer synthesis, depicted in
Figure~\ref{fig:obs_generalized_plant}, is inspired by the input-output observer
originally proposed in~\cite{marquez_2003_a_frequency_domain,
marquez_2005_robust_state_observer}.
The controller component of the observer,
$\mbs{\mathcal{K}}$ is synthesized by solving a mixed \Htwo{}-\Hinf{}~robust
control problem. The \Htwo{}~norm is a suitable performance metric for an
observer due to its interpretation as the expected RMS output of a system
subject to unit variance white noise
input~\cite[\S3.3.3]{green_linear_1994}\cite[\S5.7]{zhou_robust_1995}\cite{megretski_2004_interpretations}.
Recall~\eqref{eq:def_CG}
and let
the states and state-space matrices of
$\mbs{\mc{W}}_\mathrm{p}$,
$\mbs{\mc{W}}_\mathrm{u}$, and
$\mbs{\mc{W}}_{\!\Delta}$
be denoted similarly with superscripts.
%

A state-space realization of the generalized plant is
\begin{small}\begin{multline}
    \begin{bmatrix}
        \mbf{x}_{k+1} \\
        \mbfhat{x}_{k+1} \\
        \mbf{x}_{k+1}^{\mbs{\mathcal{W}}_\mathrm{p}} \\
        \mbf{x}_{k+1}^{\mbs{\mathcal{W}}_\mathrm{u}} \\
        \mbf{x}_{k+1}^{\mbs{\mathcal{W}}_{\!\Delta}}
    \end{bmatrix}
    =
    \begin{bmatrix}
        \mbf{A} &
        \mbf{0} &
        \mbf{0} &
        \mbf{B}\mbf{C}^{\mbs{\mc{W}}_\mathrm{u}} &
        \mbf{B}\mbf{C}^{\mbs{\mc{W}}_{\!\Delta}} \\
        \mbf{0} &
        \mbf{A} &
        \mbf{0} &
        \mbf{B}\mbf{C}^{\mbs{\mc{W}}_\mathrm{u}} &
        \mbf{0} \\
        \mbf{B}^{\mbs{\mc{W}}_\mathrm{p}} &
        -\mbf{B}^{\mbs{\mc{W}}_\mathrm{p}} &
        \mbf{A}^{\!\mbs{\mc{W}}_\mathrm{p}} &
        \mbf{0} &
        \mbf{0} \\
        \mbf{0} &
        \mbf{0} &
        \mbf{0} &
        \mbf{A}^{\!\mbs{\mc{W}}_\mathrm{u}} &
        \mbf{0} \\
        \mbf{0} &
        \mbf{0} &
        \mbf{0} &
        \mbf{0} &
        \mbf{A}^{\!\mbs{\mc{W}}_{\!\Delta}}
    \end{bmatrix}
    \begin{bmatrix}
        \mbf{x}_k \\
        \mbfhat{x}_k \\
        \mbf{x}_k^{\mbs{\mathcal{W}}_\mathrm{p}} \\
        \mbf{x}_k^{\mbs{\mathcal{W}}_\mathrm{u}} \\
        \mbf{x}_k^{\mbs{\mathcal{W}}_{\!\Delta}}
    \end{bmatrix}
    \\
    +
    \begin{bmatrix}
        \mbf{B}\mbf{D}^{\mbs{\mathcal{W}}_\mathrm{u}} &
        \mbf{B}\mbf{D}^{\mbs{\mathcal{W}}_{\!\Delta}} \\
        \mbf{B}\mbf{D}^{\mbs{\mathcal{W}}_\mathrm{u}} &
        \mbf{0} \\
        \mbf{0} &
        \mbf{0} \\
        \mbf{B}^{\mbs{\mathcal{W}}_\mathrm{u}} &
        \mbf{0} \\
        \mbf{0} &
        \mbf{B}^{\mbs{\mathcal{W}}_{\!\Delta}}
    \end{bmatrix}
    \begin{bmatrix}
        \mbf{w}_{1,k} \\
        \mbf{w}_{2,k}
    \end{bmatrix}
    +
    \begin{bmatrix}
        \mbf{0} \\
        \mbf{B} \\
        \mbf{0} \\
        \mbf{0} \\
        \mbf{0}
    \end{bmatrix}
    \mbf{u}_k,
\end{multline}
    \begin{multline}
    \begin{bmatrix}
        \mbf{z}_{1,k}\\
        \mbf{z}_{2,k}\\
    \end{bmatrix}
    =
    \begin{bmatrix}
        \mbf{D}^{\mbs{\mathcal{W}}_\mathrm{p}} &
        -\mbf{D}^{\mbs{\mathcal{W}}_\mathrm{p}} &
        \mbf{C}^{\mbs{\mathcal{W}}_\mathrm{p}} &
        \mbf{0} &
        \mbf{0} \\
        \mbf{0} &
        \mbf{0} &
        \mbf{0} &
        \mbf{C}^{\mbs{\mathcal{W}}_\mathrm{u}} &
        \mbf{C}^{\mbs{\mathcal{W}}_{\!\Delta}}
    \end{bmatrix}
    \begin{bmatrix}
        \mbf{x}_k \\
        \mbfhat{x}_k \\
        \mbf{x}_k^{\mbs{\mathcal{W}}_\mathrm{p}} \\
        \mbf{x}_k^{\mbs{\mathcal{W}}_\mathrm{u}} \\
        \mbf{x}_k^{\mbs{\mathcal{W}}_{\!\Delta}}
    \end{bmatrix}
    \\
    +
    \begin{bmatrix}
        \mbf{0} &
        \mbf{0} \\
        \mbf{D}^{\mbs{\mathcal{W}}_\mathrm{u}} &
        \mbf{D}^{\mbs{\mathcal{W}}_{\!\Delta}}
    \end{bmatrix}
    \begin{bmatrix}
        \mbf{w}_{1,k} \\
        \mbf{w}_{2,k}
    \end{bmatrix}
    +
    \begin{bmatrix}
        \mbf{0} \\
        \mbf{0}
    \end{bmatrix}
    \mbf{u}_k,
\end{multline}
    \begin{multline}
    \mbf{y}_k
    =
    \begin{bmatrix}
        \mbf{C} &
        -\mbf{C} &
        \mbf{0} &
        \mbf{0} &
        \mbf{0}
    \end{bmatrix}
    \begin{bmatrix}
        \mbf{x}_k \\
        \mbfhat{x}_k \\
        \mbf{x}_k^{\mbs{\mathcal{W}}_\mathrm{p}} \\
        \mbf{x}_k^{\mbs{\mathcal{W}}_\mathrm{u}} \\
        \mbf{x}_k^{\mbs{\mathcal{W}}_{\!\Delta}}
    \end{bmatrix}
    +
    \begin{bmatrix}
        \mbf{0} & \mbf{0}
    \end{bmatrix}
    \begin{bmatrix}
        \mbf{w}_{1,k} \\
        \mbf{w}_{2,k}
    \end{bmatrix}
    +
    \mbf{0}\,\mbf{u}_k.
\end{multline}\end{small}
Using these state-space matrices, an \Htwo{}-\Hinf{}~optimal controller is
synthesized using the approach of~\cite[\S5.4.4]{caverly_2019_lmi}.
The final observer has the structure shown in Figure~\ref{fig:obs_structure}.
\begin{figure}[htb]
    \centering
    \includegraphics[width=3.2in]{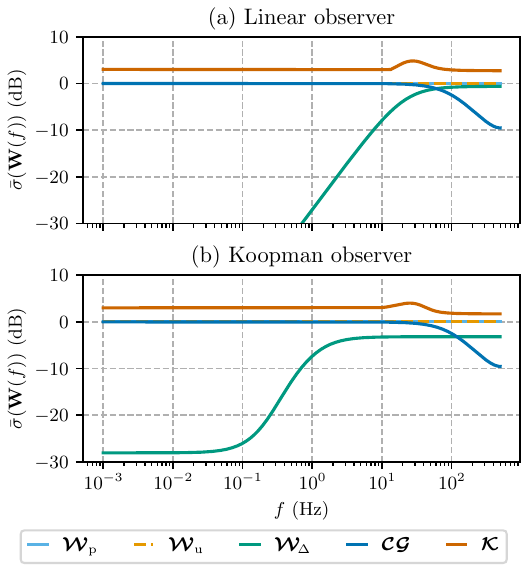}
    \caption{%
        Performance, input, and uncertainty weights for the linear (a) and
        Koopman (b) observer, along with the plant and controller frequency
        responses. The linear uncertainty weight is close to \SI{0}{\deci\bel}
        above \SI{100}{\hertz}, leading to higher controller gains and degraded
        performance at high frequencies.}\label{fig:observer_weights}
\end{figure}
\begin{figure*}[htbp]
    \centering
    \begin{subfigure}[t]{3.2in}
        \centering
        \includegraphics[width=3.2in]{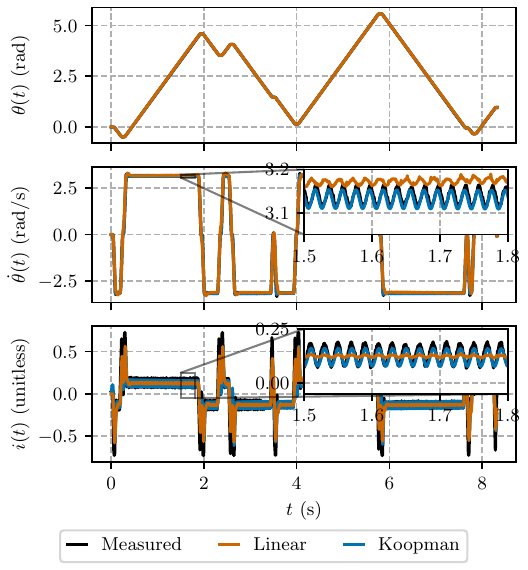}
        \caption{Nominal motor drive.}\label{fig:observer_nominal_noload_traj}
    \end{subfigure}%
    \hfill
    \begin{subfigure}[t]{3.2in}
        \centering
        \includegraphics[width=3.2in]{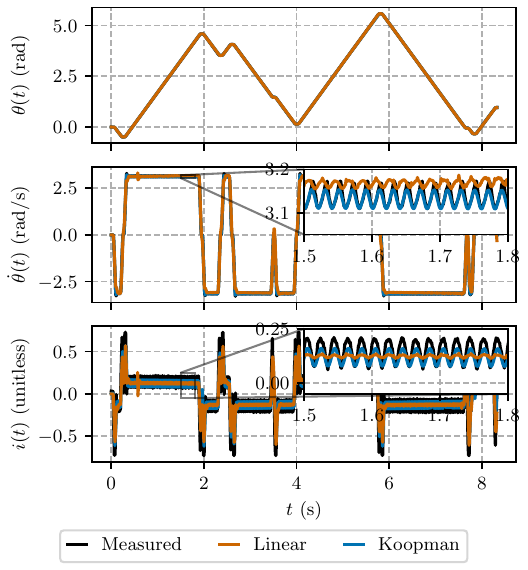}
        \caption{Off-nominal motor drive.}\label{fig:observer_offnominal_noload_traj}
    \end{subfigure}
    \caption{%
    Estimated position, velocity, and current trajectories of the linear and
    Koopman observers for the nominal and the worst off-nominal systems. The
    Koopman observer is able to accurately predict the Harmonic Drive
    oscillations while the linear observer is not.
    In fact, the linear model's transient velocity estimates are degraded due to
    higher model uncertainty at high frequencies. Off-nominal predictions are
    slightly worse for all observers.}\label{fig:observer_noload_traj}
\end{figure*}
\begin{figure*}[htbp]
    \centering
    \begin{subfigure}[t]{3.2in}
        \centering
        \includegraphics[width=3.2in]{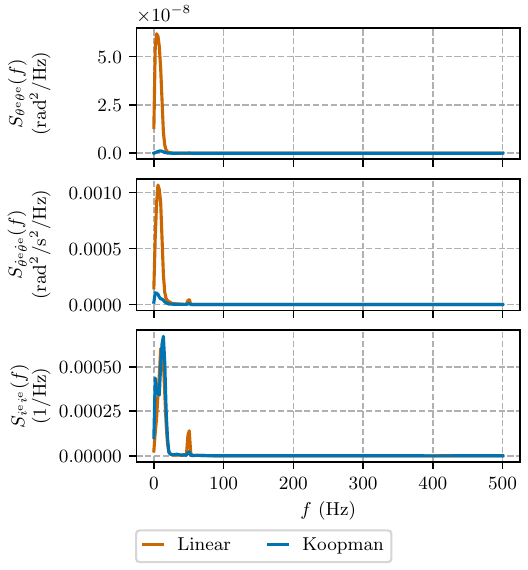}
        \caption{Nominal motor drive.}\label{fig:observer_nominal_noload_psd}
    \end{subfigure}%
    \hfill
    \begin{subfigure}[t]{3.2in}
        \centering
        \includegraphics[width=3.2in]{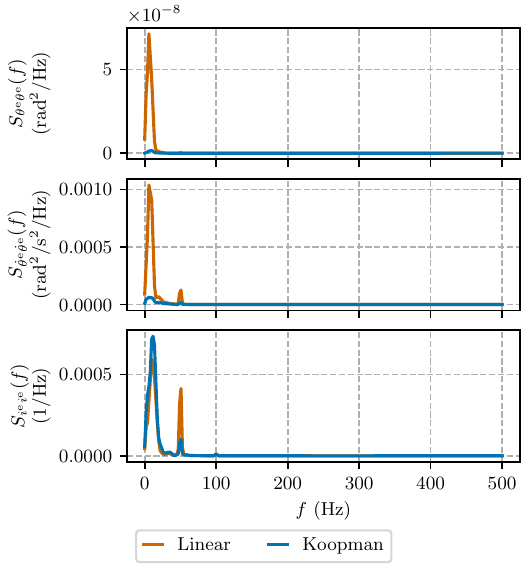}
        \caption{Off-nominal motor drive.}\label{fig:observer_offnominal_noload_psd}
    \end{subfigure}
    \caption{%
    Power spectral densities of position, velocity, and current estimation
    errors of the linear and Koopman observers for the nominal and worst
    off-nominal systems. In both cases, the Koopman observer reduces state
    estimation errors at \SI{50}{\hertz}. A linear scale is used to emphasize
    the difference in error.}\label{fig:observer_noload_psd}
\end{figure*}

The observer synthesis procedure is identical for the linear and Koopman models,
as they are both represented by state-space matrices. When using the Koopman
model for prediction in the Koopman observer, the original system states are
recovered and re-lifted with new exogenous inputs at each timestep.

\subsubsection{Observer weighting functions}
Figure~\ref{fig:observer_weights} shows the specific weighting functions used in
the robust observer synthesis problem.
The generalized plant has one performance input and one performance output:
$\mbf{w}_1$, which is the weighted system input, and $\mbf{z}_1$, which is the
weighted state estimation error.
For the sake of comparison, the input and performance weights are left as
identity. In the Koopman observer, the performance weight for the lifted state
estimation error is zero.
Only the uncertainty weight differs between the linear and Koopman
observer synthesis problems.

The intuition for how the observer weight affects the performance of the
observer is as follows.
The controller component of the observer, $\mbs{\mathcal{K}}$, governs the
degree to which the observer's internal nominal plant model is corrected.
Low uncertainty indicates that the true plant and nominal plant match well, so
the predictions of the observer's internal plant model should be trusted, and
its input should not be corrected. This corresponds to a low controller gain.
High uncertainty indicates that the true plant and the nominal plant may differ
significantly, so measurements should be used to correct the observer's internal
nominal plant model. This corresponds to a high controller gain.
In Figure~\ref{fig:observer_weights}, the controller gain rises as the model
uncertainty rises. As the plant gain rolls off, so does the controller gain. The
linear model has higher uncertainty at high frequencies, so the controller gain
remains higher than that of the Koopman model. In the next section, this effect
is shown to degrade state estimation performance.

\subsection{Experimental results}
The performance of the linear and Koopman robust observers is compared in this
section. Three test conditions are considered. First, the observers are tested
with measurements from their nominal motor drive. Then they are tested with
measurements from the furthest off-nominal motor drive. Finally, the observers
are tested with measurements from the loaded nominal motor drive.

\subsubsection{Performance with nominal and off-nominal drives}
The linear and Koopman observers are now compared for both the nominal motor
drive and the worst off-nominal motor drive.
The worst off-nominal motor drive is the motor drive with the largest residual
in Figure~\ref{fig:uncertainty_bound}.
Figure~\ref{fig:observer_noload_traj} shows the position, velocity, and current
state estimates for the nominal and off-nominal motor drives. In both cases, the
Koopman observer is able to account for the gearbox oscillation while the linear
model is not.
Due to higher model uncertainty at high frequencies, the linear observer's
transient velocity predictions are less accurate than those of the Koopman
observer.
There is a slight performance decrease going from the nominal to the off-nominal
motor drive, but the performance is still acceptable.
Figure~\ref{fig:observer_noload_psd} shows the corresponding state estimation
errors in the frequency domain. The Koopman observer significantly reduces
errors at \SI{50}{\hertz}.
\begin{figure}[htbp]
    \centering
    \includegraphics[width=3.2in]{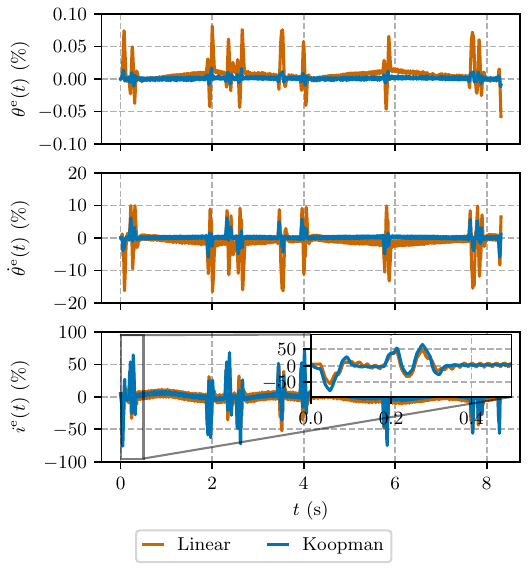}
    \caption{Position, velocity, and current estimation errors for linear and
    Koopman observers for the nominal plant with an asymmetric inertial load.
    The current estimation error $i^\mathrm{e}(t)$ can be treated as an estimate of the
    load torque. The Koopman observer's load torque estimate is less affected by
    Harmonic Drive gearbox oscillations.}\label{fig:observer_nominal_load}
\end{figure}
\begin{figure}[htbp]
    \centering
    \includegraphics[width=3.2in]{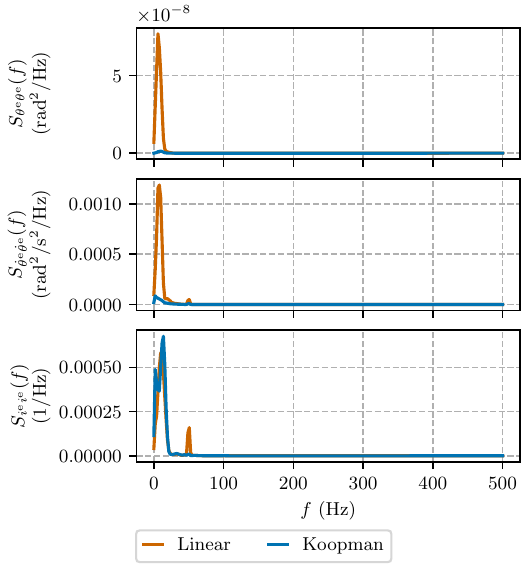}
    \caption{Power spectral densities of position, velocity, and current
    estimation errors for linear and Koopman observers for the nominal plant
    with an asymmetric inertial load. The Koopman observer's load torque
    estimate is less affected by the \SI{50}{\hertz} Harmonic Drive gearbox
    oscillations.}\label{fig:observer_nominal_load_psd}
\end{figure}
\begin{figure*}[htbp]
    \centering
    \begin{subfigure}[t]{3.2in}
        \centering
        \includegraphics[width=3.2in]{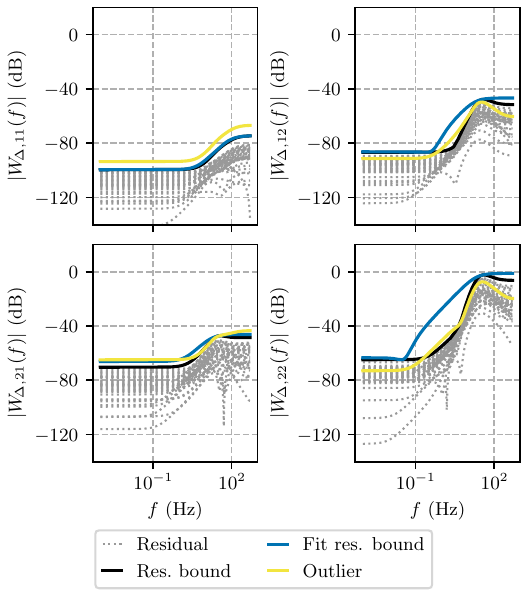}
        \caption{Linear uncertainty bounds and outliers,
        input-to-output.}\label{fig:outliers_bound_mimo_linear}
    \end{subfigure}%
    \hfill
    \begin{subfigure}[t]{3.2in}
        \centering
        \includegraphics[width=3.2in]{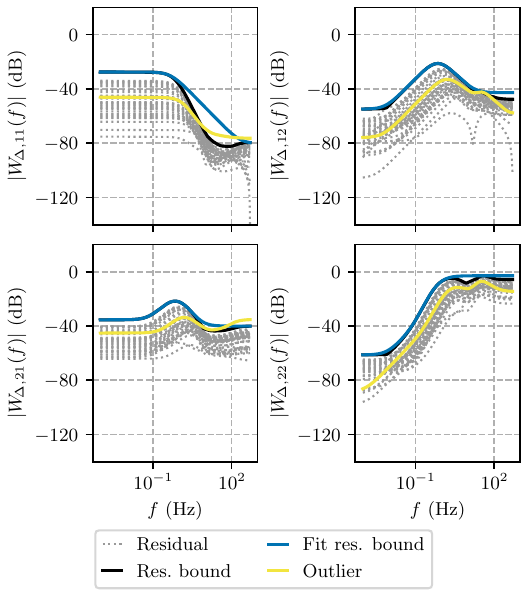}
        \caption{Koopman uncertainty bounds and outliers,
        input-to-output.}\label{fig:outliers_bound_mimo_koopman}
    \end{subfigure}
    \caption{Inverse input multiplicative uncertainty bounds, fit transfer
        functions, and outlier residual for linear and Koopman models. The
        outlier residual exceeds the fit residual bound in $W_{\Delta,11}(f)$
        and $W_{\Delta,21}(f)$ in both cases.}\label{fig:outliers_bound}
\end{figure*}

\subsubsection{Performance with loaded drive}
Observer performance is now assessed for the nominal motor drive with an
asymmetric inertial load attached. Due to the force of gravity, the load
introduces a low-frequency sinusoidal disturbance to the system.
The goal of the observer in this context is to estimate the low-frequency load
disturbance while rejecting the high-frequency gearbox disturbance.

Since the linear and Koopman models were not identified using loaded data, the
current prediction errors of their corresponding observers can be treated as
estimates of the external load torque on the motor drive, particularly when
angular acceleration is low.
The state estimation errors of these observers are shown in
Figure~\ref{fig:observer_nominal_load}, where $i^\mathrm{e}(t)$ can be treated as a
load torque estimate.
Figure~\ref{fig:observer_nominal_load_psd}, which contains the power spectral
densities of the state estimation errors, shows that the Koopman load torque
estimate contains less \SI{50}{\hertz} Harmonic Drive disturbance than the
linear load estimate.

\section{Conclusion}\label{sec:conclusion}
This paper proposes a robust nonlinear observer synthesis method based on the
Koopman operator.
Thanks to the linearity of the Koopman operator, uncertainty within a population
of Koopman models can be quantified in the frequency domain, which allows
standard robust controller synthesis tools to be used to synthesize a nonlinear
observer.
The proposed methodology is general enough that it can be applied to any system
that can be modelled with the Koopman operator.

A detailed and industrially relevant experimental example is also presented,
wherein Koopman models for a population of 38 motor drives are identified and
uncertainty within the population is quantified using standard robust control
tools. The corresponding dataset is publicly available
at~\cite{dahdah_2024_quantifying}.
%
%
Using this uncertainty model, a mixed \Htwo{}-\Hinf{}~robust observer is
designed to estimate motor velocity and current from position measurements.
While much contemporary Koopman control literature focuses on simulated systems,
the proposed observer design approach is demonstrated with real data.
Furthermore, the state-of-the-art in robust Koopman control considers only very
simple uncertainty models compared to those presented here.

As with many Koopman control approaches, one limiting factor in this approach is
that is does not lift the control inputs in the Koopman model. Bilinear lifting
approaches like~\cite{bruder_2021_advantages} show great potential for MPC
algorithms, but may be difficult to integrate with LTI controllers.
The proposed observer design approach can equally be used to synthesize robust
optimal controllers, which is a topic that will be explored in future work.

\section*{Acknowledgments}
The authors thank Martin Dionne, Eric Boutet, and Jonathan Coulombe for their
assistance in collecting the motor drive dataset used in this paper. The authors
also thank Alexandre Coulombe for providing a photo of the motor drive used in
this paper and Jonathan Eid for providing an implementation of the optimization
algorithm used to fit bounds to transfer function residuals.

\appendices

\section{Outlier Detection}\label{sec:app_outlier}
If a motor drive is installed in its frame with inappropriate fastener torques,
tracking performance is degraded and oscillations induced by the Harmonic Drive
gearbox are worsened. This effect is most apparent when the motor drive is
loaded.
The uncertainty characterization approach described in
Section~\ref{sec:experimental} can be used to identify motor drives that do not
belong to the training population, including motor drives that are installed
incorrectly.

The uncertainty characterization procedure is now repeated for the loaded
dataset, using the same optimization procedure to determine weighting functions.
The inverse input multiplicative residuals corresponding to a motor drive that
was deliberately installed with incorrect fastener torques are shown in
Figure~\ref{fig:outliers_bound}.
The incorrectly installed motor drive can be identified by looking at
$W_{\Delta,11}(f)$ in Figure~\ref{fig:outliers_bound_mimo_linear} and
$W_{\Delta,21}(f)$ in Figure~\ref{fig:outliers_bound_mimo_koopman}, where the
outlier residuals exceed the fit residual bound.

While inverse input multiplicative uncertainty is shown in this section, the
outlier drive is detectable using any uncertainty model.
It has been verified that this outlier identification criterion is insensitive
to the choice of nominal model. All choices of nominal model except one
correctly identify the outlier model. It could be argued that that nominal model
should also be considered an outlier.
Ultimately, the decision of whether a system should be considered an inlier or
outlier depends on the use case. The uncertainty mode could equally be extended
to include the incorrectly installed drive if a more conservative uncertainty
model is acceptable.

\section{Harmonic Drive Phase Calibration}\label{sec:app_calibration}
The Koopman lifting functions in~\eqref{eq:state_dependent_lifting} include a
fixed parameter for the phase of the nonlinear oscillation term, which must be
determined before computing the Koopman matrix.
Since this parameter is determined when the motor drive is being assembled, it
is unknown but constant.
This section proposes a calibration procedure for this unknown phase offset that
can easily be incorporated into the motor drive's existing calibration
routine.

Designing lifting functions to include both $\cos{(100\,\theta_k)}$ and
$\sin{(100\,\theta_k)}$ could account for an unknown phase offset by absorbing
it into the Koopman matrix.
While this may be appropriate when identifying a single motor drive, it
significantly and artificially increases uncertainty within a population of
motor drives, negatively impacting observer performance.
In an earlier version of this work, the phase offset is determined through a
hyperparameter optimization procedure~\cite{dahdah_closed-loop_v1_2023}.
However, a simpler and more reliable method is presented in this paper.

To find the phase offset for a given drive, the position and velocity
trajectories are separated into constant-velocity segments.
For each segment $i$, the velocity tracking error $\dot{\theta}_i^\mathrm{e}$ is
calculated and normalized, and the optimum phase,
\begin{equation}
    \varphi_i = \arg\max_{\hat{\varphi}_i}\left\langle
    \dot{\theta}_i^\mathrm{e},\,\sin(100\,\theta_i + \hat{\varphi_i})
    \right\rangle,
\end{equation}
is found by evaluating 1000 phase samples in $[0, 2\pi)$.
The final phase offset is then calculated by averaging the optimal phase of each
segment using the circular mean~\cite[\S2.2.1]{mardia_1999_directional},
\begin{equation}
    \varphi
    =
    \mathrm{atan2}\left(%
        \frac{1}{N} \sum_{i=1}^N \sin \varphi_i,
        \frac{1}{N} \sum_{i=1}^N \cos \varphi_i
    \right),
\end{equation}
where $N$ is the number of segments.
To avoid this offline calibration procedure, a phase-locked loop could be used
to estimate $\varphi$ online from velocity tracking errors.

\bibliographystyle{IEEEtran}
\bibliography{paper}

\begin{IEEEbiography}[{\includegraphics[width=1in,height=1.25in,clip,keepaspectratio]{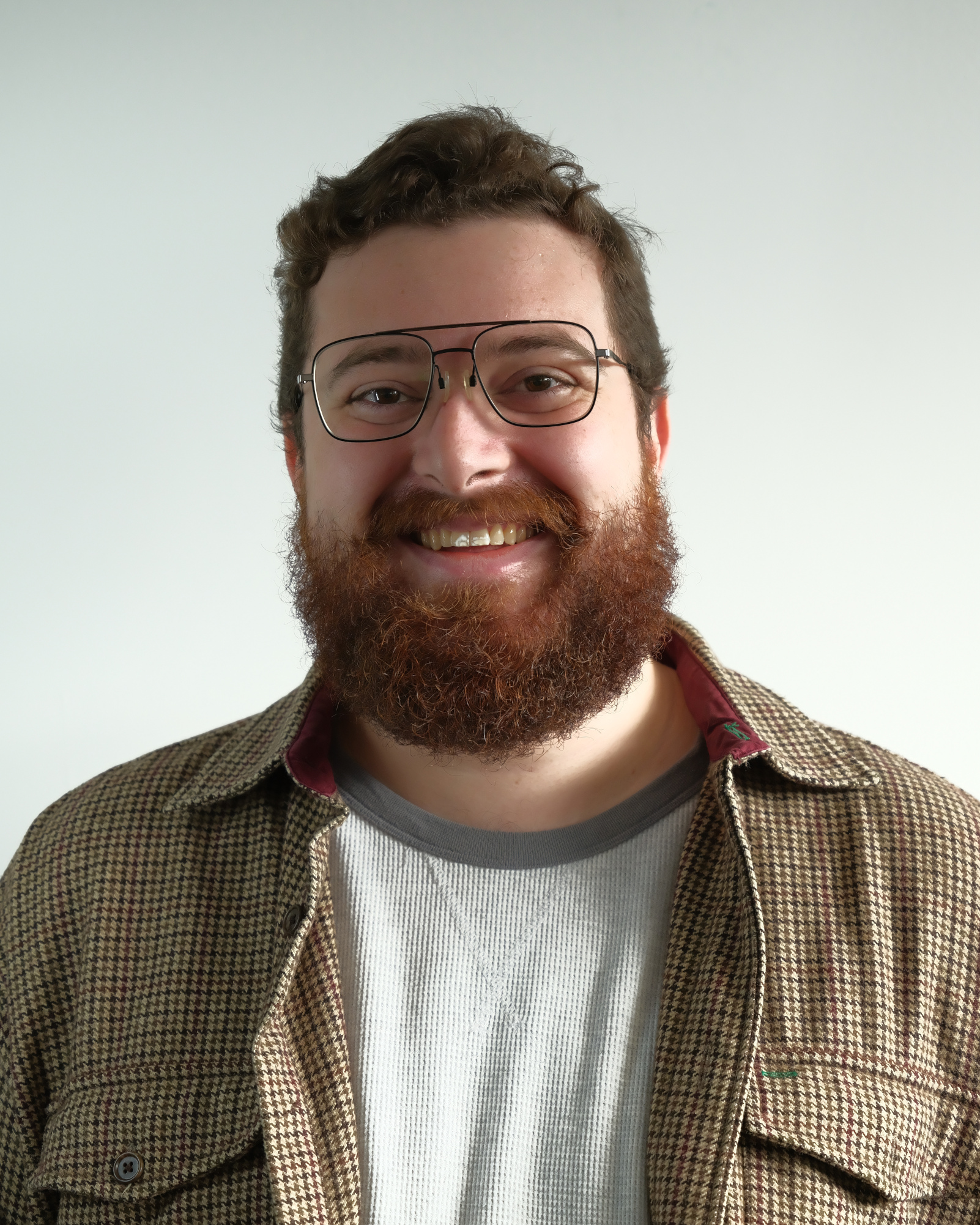}}]{%
        Steven Dahdah} (Student Member, IEEE)
    received the B.Eng.\ degree in electrical engineering (Honours, Co-op)
    with a minor in software engineering from McGill University, Montreal
    QC, Canada in 2019.

    He is currently a Ph.D.\ candidate in the Department of Mechanical
    Engineering at McGill university. His research interests include
    control systems, system identification, and machine learning.
\end{IEEEbiography}

\begin{IEEEbiography}[{\includegraphics[width=1in,height=1.25in,clip,keepaspectratio]{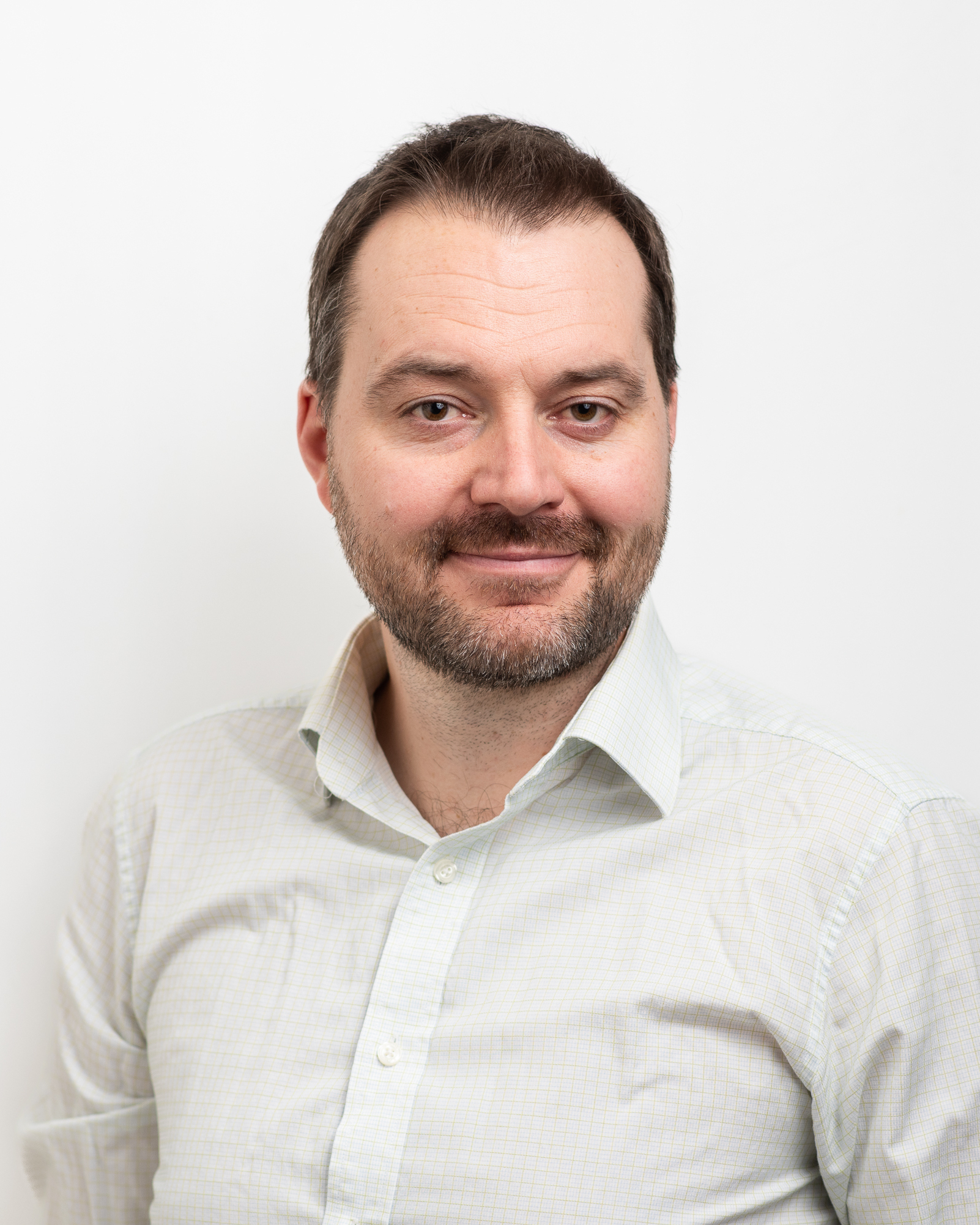}}]{%
        James Richard Forbes}
    (Member, IEEE) received the B.A.Sc.\ degree in
    mechanical engineering (Honours, Co-op) from the University of Waterloo,
    Waterloo ON, Canada, in 2006, and the M.A.Sc.\ and Ph.D.\ degrees in
    aerospace science and engineering from the University of Toronto Institute
    for Aerospace Studies, Toronto ON, in 2008 and 2011, respectively.

    He is currently an Associate Professor and William Dawson Scholar with the
    Department of Mechanical Engineering, McGill University, Montreal QC,
    Canada. His research interests include navigation, guidance, and control of
    robotic systems.

    Dr.\ Forbes is a Member of the Centre for Intelligent Machines, a Member of
    the Group for Research in Decision Analysis, and a Member of the Trottier
    Institute for Sustainability in Engineering and Design. He was the recipient
    of the McGill Association of Mechanical Engineers Professor of the Year
    Award in 2016, the Engineering Class of 1944 Outstanding Teaching Award in
    2018, the Carrie M.\ Derick Award for Graduate Supervision and Teaching in
    2020, and the Samuel and Ida Fromson Outstanding Teaching Award in
    Engineering in 2024. He is currently a Senior Editor for the International
    Journal of Robotics Research (IJRR).
\end{IEEEbiography}

\end{document}